\documentclass[aps,prb,twocolumn,showpacs,preprintnumbers,amsmath,amssymb]{revtex4}

\usepackage{graphicx}% Include figure files
\usepackage{dcolumn}% Align table columns on decimal point
\usepackage{bm}% bold math

\begin{document}

\title{Mechanisms for the decomposition and dehydrogenation of Li amide/imide}% \\

\author{Khang Hoang}
\altaffiliation{Current address: Naval Research Laboratory, Washington, D.C. 20375, USA, and George Mason University, Fairfax, VA 22030, USA. E-mail: hoang@dave.nrl.navy.mil.}%
\author{Anderson Janotti}
\author{Chris G. Van de Walle}
\email{vandewalle@mrl.ucsb.edu}
\affiliation{Materials Department, University of California, Santa Barbara, California 93106-5050, USA}%

%\date{\today}

\begin{abstract}
Reversible reaction involving Li amide (LiNH$_{2}$) and Li imide (Li$_{2}$NH) is a potential mechanism for hydrogen storage. Recent synchrotron x-ray diffraction experiments [W. I. David {\it et al.}, J. Am. Chem. Soc. {\bf 129}, 1594 (2007)] suggest that the transformation between LiNH$_{2}$ and Li$_{2}$NH is a bulk reaction that occurs through non-stoichiometric processes and involves the migration of Li$^{+}$ and H$^{+}$ ions. In order to understand the atomistic mechanisms behind these processes, we carry out comprehensive first-principles studies of native point defects and defect complexes in the two compounds. We find that both LiNH$_{2}$ and Li$_{2}$NH are prone to Frenkel disorder on the Li sublattice. Lithium interstitials and vacancies have low formation energies and are highly mobile, and therefore play an important role in mass transport and ionic conduction. Hydrogen interstitials and vacancies, on the other hand, are responsible for forming and breaking N$-$H bonds, which is essential in the Li amide/imide reaction. Based on the structure, energetics, and migration of hydrogen-, lithium-, and nitrogen-related defects, we propose that LiNH$_{2}$ decomposes into Li$_{2}$NH and NH$_{3}$ according to two competing mechanisms with different activation energies: one mechanism involves the formation of native defects in the interior of the material, the other at the surface. As a result, the prevailing mechanism and hence the effective activation energy for decomposition depend on the surface-to-volume ratio or the specific surface area, which changes with particle size during ball milling. These mechanisms also provide an explanation for the dehydrogenation of LiNH$_{2}$+LiH mixtures.

\end{abstract}


\maketitle

\section{\label{sec:intro}Introduction}

Hydrogen is a promising energy carrier in future energy systems, but storage of hydrogen is still a major challenge.\cite{eberle}
Lithium amide (LiNH$_{2}$) is a promising material due to its high hydrogen density. Lithium imide (Li$_{2}$NH) is known for its high ionic conductivity (3$\times$10$^{-4}$ S/cm at 25$^{\circ}$C).~\cite{boukamp} These two compounds have attracted a lot of attention ever since Chen {\it et al.}\cite{chenNATURE} demonstrated that Li$_{3}$N can absorb/desorb hydrogen at reasonable pressures following the reversible reaction:
\begin{equation}\label{eq:reaction1}
\mathrm{Li}_{3}\mathrm{N} + 2\mathrm{H}_{2} \leftrightarrow \mathrm{Li}_{2}\mathrm{NH} + \mathrm{LiH} + \mathrm{H}_{2} \leftrightarrow \mathrm{LiNH}_{2} + 2\mathrm{LiH}.
\end{equation}
The theoretical amount of reversible hydrogen storage in this reaction is $\sim$11.5 wt\% (expressed per mole of Li$_{3}$N). At temperatures below 300$^{\circ}$C, LiNH$_{2}$ was observed to reversibly store $\sim$6.5 wt\% hydrogen during desorption and absorption under 0.04 and 20 bar, respectively, following the reaction:~\cite{chenNATURE}
\begin{equation}\label{eq:reaction2}
\mathrm{LiNH}_{2} + \mathrm{LiH} \leftrightarrow \mathrm{Li}_{2}\mathrm{NH} + \mathrm{H}_{2}.
\end{equation}
The drawback of this Li amide/imide reaction is that the dehydrogenation temperature and hydrogenation pressure are relatively high for practical applications. Yet, the fundamental mechanisms behind the decomposition and (de)hydrogenation processes are not fully understood, and we expect that once such understanding has been established, one can provide solutions for speeding up the reaction kinetics and lowering the dehydrogenation temperature and hydrogenation pressure.

Regarding the dehydrogenation reaction in Eq.~(\ref{eq:reaction2}), it has been suggested that LiNH$_{2}$ may react directly with LiH at the LiNH$_{2}$/LiH interface according to a polar mechanism to produce H$_{2}$.~\cite{chenNATURE,chenJPCB,luIC} The mechanism is explained in terms of the strong affinity between protonic hydrogen (H$^{\delta+}$) in LiNH$_{2}$ and hydridic hydrogen (H$^{\delta-}$) in LiH where the redox reaction of H$^{\delta+}$ and H$^{\delta-}$ produces molecular hydrogen (H$_{2}$).~\cite{chenJPCB} Thermal desorption measurements carried out on a LiNH$_{2}$+2LiD mixture, however, showed that it produces mainly H$_{2}$ in addition to HD and D$_{2}$ (instead of mainly HD as one would have expected).~\cite{chenJPCB} This seems to be contrary to the redox hypothesis.

Others have proposed that NH$_{3}$ necessarily evolves as a transient gas and the dehydrogenation of LiNH$_{2}$+LiH mixtures involves an intermediate step:~\cite{huIECR,huJPCA,ichikawa04,ichikawaJPCB,pinkerton05,meisner,ichikawa05,isobe}
\begin{equation}\label{eq:reaction3}
2\mathrm{LiNH}_{2} \rightarrow \mathrm{Li}_{2}\mathrm{NH} + \mathrm{NH}_{3};
\end{equation}
\begin{equation}\label{eq:reaction4}
\mathrm{NH}_{3} + \mathrm{LiH} \rightarrow \mathrm{LiNH}_{2} + \mathrm{H}_{2}.
\end{equation}
The first reaction releases 37 wt\% NH$_{3}$ and was suggested to be diffusion-controlled, whereas the second reaction releases 5.8 wt\% H$_{2}$ and is supposedly ultrafast. The decomposition of LiNH$_{2}$ into Li$_{2}$NH and NH$_{3}$ is well known,~\cite{huJPCA,chenJPCB,ichikawa04} and it was Hu and Ruckenstein who pointed out that NH$_{3}$ reacts quickly with LiH.~\cite{huIECR,huJPCA} The activation energy for the decomposition of LiNH$_{2}$ was estimated to be 2.53 eV (before ball milling), and it was found to decrease with increasing ball-milling time.~\cite{markmaitree} The above two-step pathway is supported by recent studies using variable-temperature {\it in situ} $^\mathrm{1}$H NMR spectroscopy.~\cite{huJPS}

As noted by David {\it et al.},\cite{davidJACS} there are very close structural similarities between the tetragonal LiNH$_{2}$ and the antifluorite Li$_{2}$NH. Through structural refinement from synchrotron x-ray diffraction data, they suggested that the transformation between LiNH$_{2}$ and Li$_{2}$NH is a bulk reaction that occurs through non-stoichiometric processes within the cubic Li-N-H structure. David {\it et al.}~further proposed a mechanism for the Li amide/imide decomposition and hydrogenation processes (within the abovementioned ammonia-mediated two-step reaction) that involves the migration of both Li$^{+}$ and H$^{+}$ ions; they also suggested that the non-stoichiometry observed in the Li-N-H system is a direct result of the ionic mobility. The most important step in this mechanism would be the movement of a lithium ion to an interstitial site, forming a lithium Frenkel defect pair.~\cite{davidJACS}

In addition to the polar mechanism and the ammonia-mediated mechanism, Aguey-Zinsou {\it et al.}\cite{Aguey-Zinsou} have recently suggested that the reaction between LiNH$_{2}$ and LiH below 300$^{\circ}$C is a heterogeneous solid-state reaction, controlled by the diffusion of Li$^{+}$ from LiH to LiNH$_{2}$ across the interface. In this mechanism, the reaction is direct rather than ammonia-mediated.~\cite{Aguey-Zinsou}

Theoretical studies of LiNH$_{2}$ and Li$_{2}$NH to date have focused mainly on structural, electronic, and thermodynamic properties of the bulk compounds.~\cite{herbst,miwa,yangAPL,song,mueller,kope,tsumuraya}  Experimental data,~\cite{davidJACS} on the other hand, suggest that the rate-limiting process in the Li amide/imide reaction involves mass transport mediated by point defects. This scenario motivated us to perform first-principles calculations for point defects and defect complexes in LiNH$_{2}$ and Li$_{2}$NH in order to explore possible defect-related mechanisms that can explain the decomposition of LiNH$_{2}$ [reaction (\ref{eq:reaction3})] and the hydrogenation of Li$_{2}$NH.  Some preliminary results and partial conclusions of our work have been reported elsewhere.~\cite{hoang_angew} Other research groups have also recently started investigating native defects,~\cite{miceli,hazrati,wang} but our study goes much further in identifying specific mechanisms that can explain the experimental observations.  A detailed comparison with the previous papers will be addressed in Sections \ref{ssec:N} and \ref{ssec:decomp}.

Indeed, we show that LiNH$_{2}$ decomposes into Li$_{2}$NH and NH$_{3}$ via two competing mechanisms with different activation energies: one mechanism involves the formation of native defects in the interior of the material, the other at the surface. As a result, the prevailing mechanism and hence the effective activation energy for decomposition depend on the surface-to-volume ratio or the specific surface area, which changes with particle size during ball milling. The dehydrogenation of LiNH$_{2}$+LiH mixtures can be explained in terms of the two-step reaction [Eqs.~(\ref{eq:reaction3}) and (\ref{eq:reaction4})] and the mechanisms we propose for LiNH$_{2}$ decomposition. However, NH$_{3}$ is not necessarily formed and released from a LiNH$_{2}$+LiH mixture if LiNH$_{2}$ and LiH are in intimate contact.

We also show that lithium interstitials and vacancies in LiNH$_{2}$ and Li$_{2}$NH can be formed in the interior of the materials via a Frenkel-pair mechanism and are highly mobile, and that Li amide (imide) units can be locally formed inside the bulk Li imide (amide). Our results support David {\it et al.}'s proposal that the Li amide/imide is a bulk reaction, and that there is a continuous transformation between LiNH$_{2}$ and Li$_{2}$NH via non-stoichiometric intermediates.~\cite{davidJACS} It is, however, not the formation and migration of lithium-related defects that is the rate-limiting step in the kinetics of the Li amide/imide reaction, but the formation and migration of hydrogen interstitials and vacancies which are responsible for forming and breaking N$-$H bonds in LiNH$_{2}$ (and Li$_{2}$NH).

The remainder of this paper is arranged as follows: in Sec.~II we provide technical details of the calculations and present the theoretical approach. Bulk properties of LiNH$_{2}$ and Li$_{2}$NH are discussed in Sec.~III. In Secs.~IV and V, we present the results for native defects and discuss their relevance to ionic conduction in LiNH$_{2}$ and Li$_{2}$NH, decomposition of LiNH$_{2}$, dehydrogenation of LiNH$_{2}$+LiH mixtures, and hydrogenation of Li$_{2}$NH. A summary in Sec.~VI concludes the paper.

\section{\label{sec:metho}Methodology}

\subsection{Computational details}

Our calculations were based on density-functional theory within the generalized-gradient approximation (GGA)~\cite{GGA} and the projector augmented wave method,~\cite{PAW1,PAW2} as implemented in the VASP code.~\cite{VASP1,VASP2,VASP3} Calculations for bulk LiNH$_{2}$ (tetragonal $I\overline{4}$; 32 atoms/unit cell) were performed using a 10$\times$10$\times$5 Monkhorst-Pack $\mathbf{k}$-point mesh;~\cite{monkhorst-pack} for Li$_{2}$NH (orthorhombic $Pbca$; 32 atoms/unit cell) we used a 10$\times$5$\times$10 $\mathbf{k}$-point mesh. For defect calculations, we used a (2$\times$2$\times$1) supercell for LiNH$_{2}$ and a (2$\times$1$\times$2) supercell for Li$_{2}$NH, both corresponding to 128 atoms/cell, and a 2$\times$2$\times$2 $\mathbf{k}$-point mesh and plane-wave basis-set cutoff of 400 eV. In these calculations, the lattice parameters were fixed to the calculated bulk values, but all the internal coordinates were fully relaxed. Convergence with respect to self-consistent iterations was assumed when the total energy difference between cycles was less than 10$^{-4}$ eV and the residual forces were better than 0.01 eV/{\AA}. The migration of selected native point defects in LiNH$_{2}$ and Li$_{2}$NH was studied using the climbing image nudged elastic band method (NEB).~\cite{ci-neb}

\subsection{Defect formation energies}

Throughout the paper we will use defect formation energies to characterize different native defects in LiNH$_{2}$ and Li$_{2}$NH. The formation energy ($E^{f}$) of a defect is a crucial factor in determining its concentration. In thermal equilibrium, the concentration of the defect X at temperature $T$ can be obtained via the relation~\cite{vdwJAP,janotti2009}
\begin{equation}\label{eq:concen}
c(\mathrm{X})=N_{\mathrm{sites}}N_{\mathrm{config}}\mathrm{exp}[-E^{f}(\mathrm{X})/k_BT],
\end{equation}
where $N_{\mathrm{sites}}$ is the number of high-symmetry sites in the lattice per unit volume on which the defect can be incorporated, and $N_{\mathrm{config}}$ is the number of equivalent configurations (per site). Note that the energy in Eq.~(\ref{eq:concen}) is, in principle, a free energy; however, the entropy and volume terms are often neglected because they are negligible at relevant experimental conditions.\cite{janotti2009} It emerges from Eq.~(\ref{eq:concen}) that defects with low formation energies will easily form and occur in high concentrations.

The formation energy of a defect X in charge state $q$ is defined as~\cite{vdwJAP,peles07}
\begin{eqnarray}\label{eq:eform}
\nonumber
E^f({\mathrm{X}}^q)=E_{\mathrm{tot}}({\mathrm{X}}^q)&-&E_{\mathrm{tot}}({\mathrm{bulk}})-\sum_{i}{n_i\mu_i} \\
&+&q(E_{\mathrm{v}}+\Delta V+\mu_{e}),
\end{eqnarray}
where $E_{\mathrm{tot}}(\mathrm{X}^{q})$ and $E_{\mathrm{tot}}(\mathrm{bulk})$ are, respectively, the total energies of a supercell containing the defect X and of a supercell of the perfect bulk material; $\mu_{i}$ is the atomic chemical potential of species $i$ (referenced to the standard state), and $n_{i}$ denotes the number of atoms of species $i$ that have been added ($n_{i}$$>$0) or removed ($n_{i}$$<$0) to form the defect. $\mu_{e}$ is the electron chemical potential, i.e., the Fermi level, referenced to the valence-band maximum in the bulk ($E_{\mathrm{v}}$). $\Delta V$ is the ``potential alignment'' term, i.e., the shift in the band positions due to the presence of the charged defect and the neutralizing background, obtained by aligning the average electrostatic potential in regions far away from the defect to the bulk value.~\cite{vdwJAP}

\subsection{Chemical potentials}

We note that the atomic chemical potentials $\mu_{i}$ are variables and can be chosen to represent experimental conditions. Given the reported continuous transformation between LiNH$_{2}$ and Li$_{2}$NH,\cite{davidJACS} for reactions (\ref{eq:reaction2}) and (3), it is reasonable to assume that the two compounds are in equilibrium; i.e., the chemical potentials simultaneously satisfy:
\begin{equation}\label{eq:LiNH2}
\mu_{\rm Li} + \mu_{\rm N} + 2 \mu_{\rm H} = \Delta H_f({\rm LiNH_{2}}),
\end{equation}
\begin{equation}\label{eq:Li2NH}
2\mu_{\rm Li} + \mu_{\rm N} + \mu_{\rm H} = \Delta H_f({\rm Li_{2}NH});
\end{equation}
where $\Delta H_f$ is the enthalpy of formation. The calculated formation enthalpies (at $T$=0 K) are $-$2.065 eV and $-$2.091 eV for LiNH$_{2}$ and Li$_{2}$NH, respectively, in good agreement with previously reported values.~\cite{herbst,siegelPRB,herbstAPL,mueller}

From Eqs.~(\ref{eq:LiNH2}) and (\ref{eq:Li2NH}), the chemical potentials of Li and N can be expressed in terms of $\mu_{\rm H}$, which is now the only variable. The temperature and pressure values at which the dehydrogenation and hydrogenation processes occur then determine the chemical potential of H through equilibrium with H$_2$ gas. In the following discussion, we employ a set of conditions used by David {\it et al.}~in their experiments, i.e., we use 10$^{-3}$ bar and 260$^{\circ}$C for hydrogen desorption, and 3 bar and 260$^{\circ}$C for absorption.~\cite{davidJACS} These conditions correspond to $\mu_{\rm H}$=$-$0.49 eV and $\mu_{\rm H}$=$-$0.31 eV, respectively.~\cite{H2gas} Two different sets of experimental conditions will be analyzed. $\mu_{\rm H}$=$-$0.49 eV corresponds to the dehydrogenation process and is therefore appropriate for analysis of defects in LiNH$_{2}$. $\mu_{\rm H}$=$-$0.31 eV, on the other hand, corresponds to the hydrogen absorption process, and is therefore the value we will use for analysis of defects in Li$_{2}$NH.

One can, of course, choose a different set of atomic chemical potentials which corresponds to different experimental conditions, and this may affect the relative formation energy between different defects. These formation energies can easily be obtained from the data we report. However, we have checked that the details of the choice we made here do not affect the physics of the mechanisms we are presenting.

\section{\label{sec:bulk}Bulk Properties}

LiNH$_{2}$ was reported to crystallize in the tetragonal space group $I\overline{4}$.~\cite{yangAPL} The crystal structure of Li$_{2}$NH was, however, difficult to resolve. Using x-ray diffraction, Juza and Opp proposed that Li$_{2}$NH had an antifluorite structure with the $Fm\overline{\mathrm{3}}m$ symmetry,~\cite{juza} but they were unable to obtain the positions of the hydrogen ions. More recent experimental studies suggested that hydrogen randomly occupies one of the sites around the nitrogen ion.~\cite{noritake,ohoyama}

On the theory side, significant efforts have been focused on finding low-energy ordered structures for Li$_{2}$NH and several structural models have been proposed.~\cite{herbst,kope,mueller} Among these models, the orthorhombic structure with the $Pbca$ symmetry proposed by Mueller and Ceder was shown to have the lowest energy.~\cite{mueller} We therefore employ this structure for our current studies of Li$_{2}$NH.

\begin{figure}
\begin{center}
\includegraphics[width=3.0in]{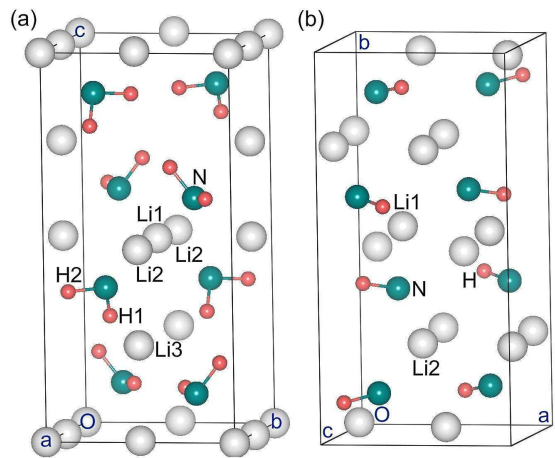}
\end{center}
\vspace{-0.15in}
\caption{(Color online) Relaxed structures of (a) tetragonal LiNH$_{2}$ and (b) orthorhombic Li$_{2}$NH. Large (gray) spheres are Li, medium (blue) spheres N, and small (red) spheres H. Inequivalent atoms are labeled as H1, H2, Li1, Li2, and Li3.}\label{structures}
\end{figure}

The optimized structures of LiNH$_{2}$ and Li$_{2}$NH are shown in Figs.~\ref{structures}(a) and \ref{structures}(b). For LiNH$_{2}$, the calculated lattice parameters are $a$=$b$=5.053 {\AA}, and $c$=10.304 {\AA}, in satisfactory agreement with experimental values ($a$=$b$=5.034 {\AA}, $c$=10.256 {\AA}).~\cite{yangAPL} For Li$_{2}$NH, we find $a$=5.134 {\AA}, $b$=10.461 {\AA}, and $c$=5.28 {\AA}, in good agreement with the values reported by Mueller and Ceder.~\cite{mueller}

We can consider the bonding in LiNH$_{2}$ as composed of (Li)$^+$ and (NH$_{2}$)$^-$ units, like the ionic bonding in NaCl; the (NH$_{2}$)$^-$ units are surrounded by (Li)$^+$ and vice versa. Similarly, Li$_{2}$NH can be regarded as composed of (Li)$^+$ and (NH)$^{2-}$ units, where for each (NH)$^{2-}$ unit there are two (Li)$^+$ units. This picture will be useful when we discuss the energetics and local geometry of various defects in LiNH$_{2}$ and Li$_{2}$NH.

\begin{figure}
\begin{center}
\includegraphics[width=2.9in]{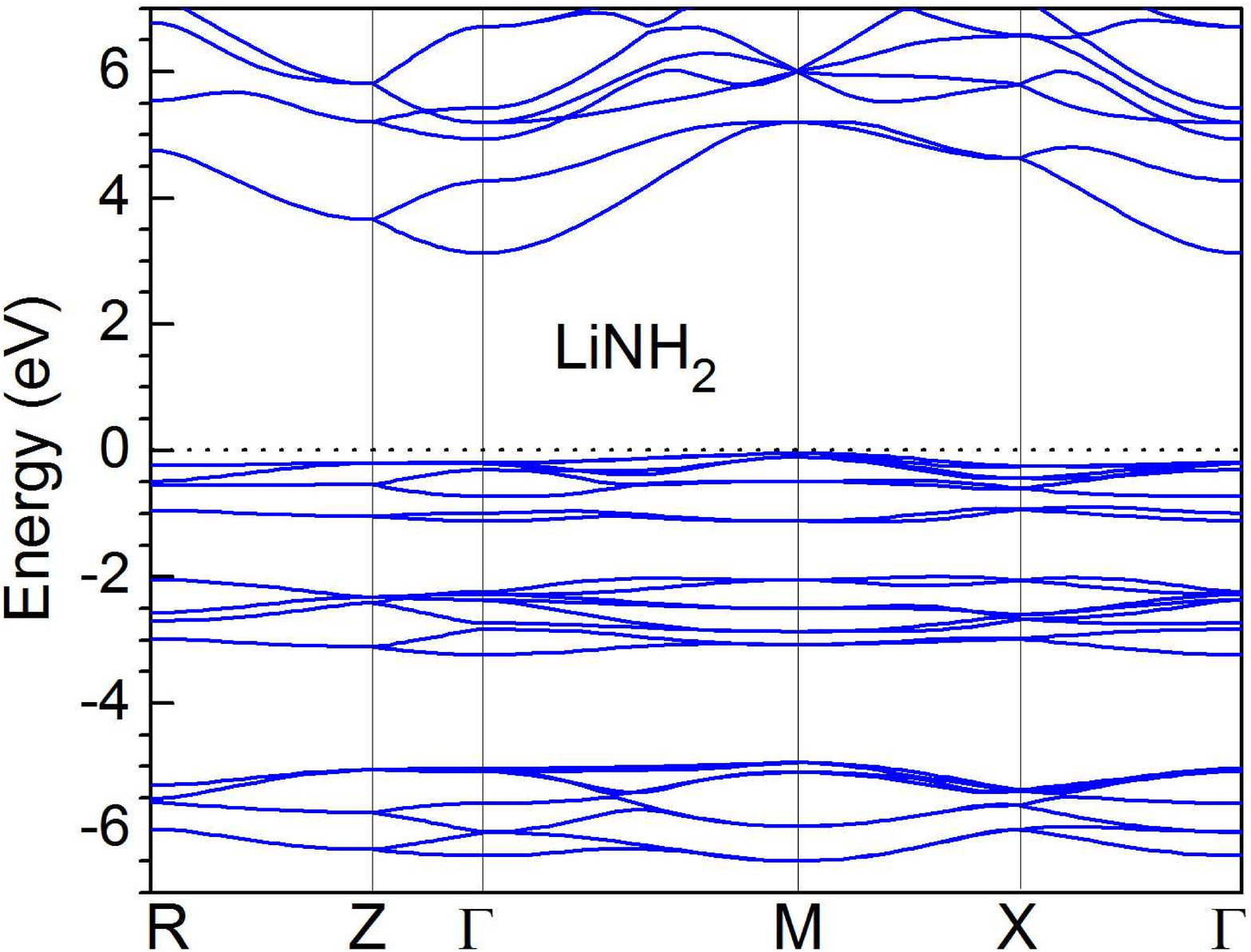}
\end{center}
\vspace{-0.15in}
\caption{(Color online) Band structure of tetragonal LiNH$_{2}$ along the high symmetry directions of the tetragonal BZ. The VBM is at the $M$ point, whereas the CBM is at the $\Gamma$ point. The zero of energy is set to the highest occupied state.}\label{LiNH2;Bands}
\end{figure}

Figure~\ref{LiNH2;Bands} shows the calculated band structure of tetragonal LiNH$_{2}$ along the high-symmetry directions of the Brillouin zone (BZ). We find an indirect band gap of 3.17 eV with the valence-band maximum (VBM) at the $M$ point and the conduction-band minimum (CBM) at the $\Gamma$ point. Band-gap values ranging from $\sim$3 to 3.48 eV have been reported for LiNH$_{2}$.~\cite{herbst,miwa,yangAPL} An analysis of the wavefunctions shows that the VBM is composed of N-related unbonded states from the (NH$_{2}$)$^-$ units, whereas the CBM is composed of a mixture of N $p$ and H $s$ states.

\begin{figure}
\begin{center}
\includegraphics[width=2.9in]{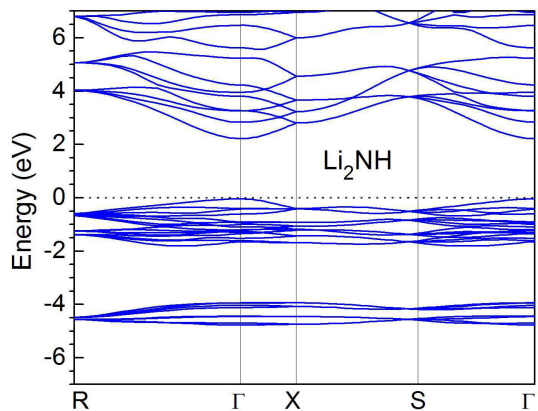}
\end{center}
\vspace{-0.15in}
\caption{(Color online) Band structure of orthorhombic Li$_{2}$NH along the high symmetry directions of the orthorhombic BZ. The VBM and CBM are at the $\Gamma$ point. The zero of energy is set to the highest occupied state.}\label{Li2NH;Bands}
\end{figure}

Figure~\ref{Li2NH;Bands} shows the calculated band structure of orthorhombic Li$_{2}$NH along the high-symmetry directions of the orthorhombic BZ. We find a direct band gap of 2.26 eV at the $\Gamma$ point. Similar to LiNH$_{2}$, the VBM of Li$_{2}$NH is composed mostly of N-related unbonded states from the (NH)$^{2-}$ units, whereas the CBM is composed of N $p$ and H $s$ states. Previous studies reported a band gap of 2.65 eV for Li$_{2}$NH.~\cite{tsumuraya} To the best of our knowledge, no experimental information on the band gaps of LiNH$_{2}$ and Li$_{2}$NH is available. As we illustrate in Sec.~IV, knowing the nature of the electronic states near the VBM and CBM is extremely helpful in understanding the formation of defects in these systems.

\section{\label{sec:defects}Point Defects and Complexes}

We investigated hydrogen-, lithium-, and nitrogen-related point defects in all the possible charge states in LiNH$_{2}$ and Li$_{2}$NH. Defect complexes were also considered, with special attention to Frenkel pairs, i.e., interstitial-vacancy pairs of the same species. Defect formation energies and migration barriers were obtained using the methods described in Sec.~\ref{sec:metho}. We also discuss the role of these native defects in mass transport and ionic conduction in LiNH$_{2}$ and Li$_{2}$NH.

\subsection{LiNH$_{2}$}

\subsubsection{Hydrogen-related defects}

\begin{figure}
\begin{center}
\includegraphics[width=3.0in]{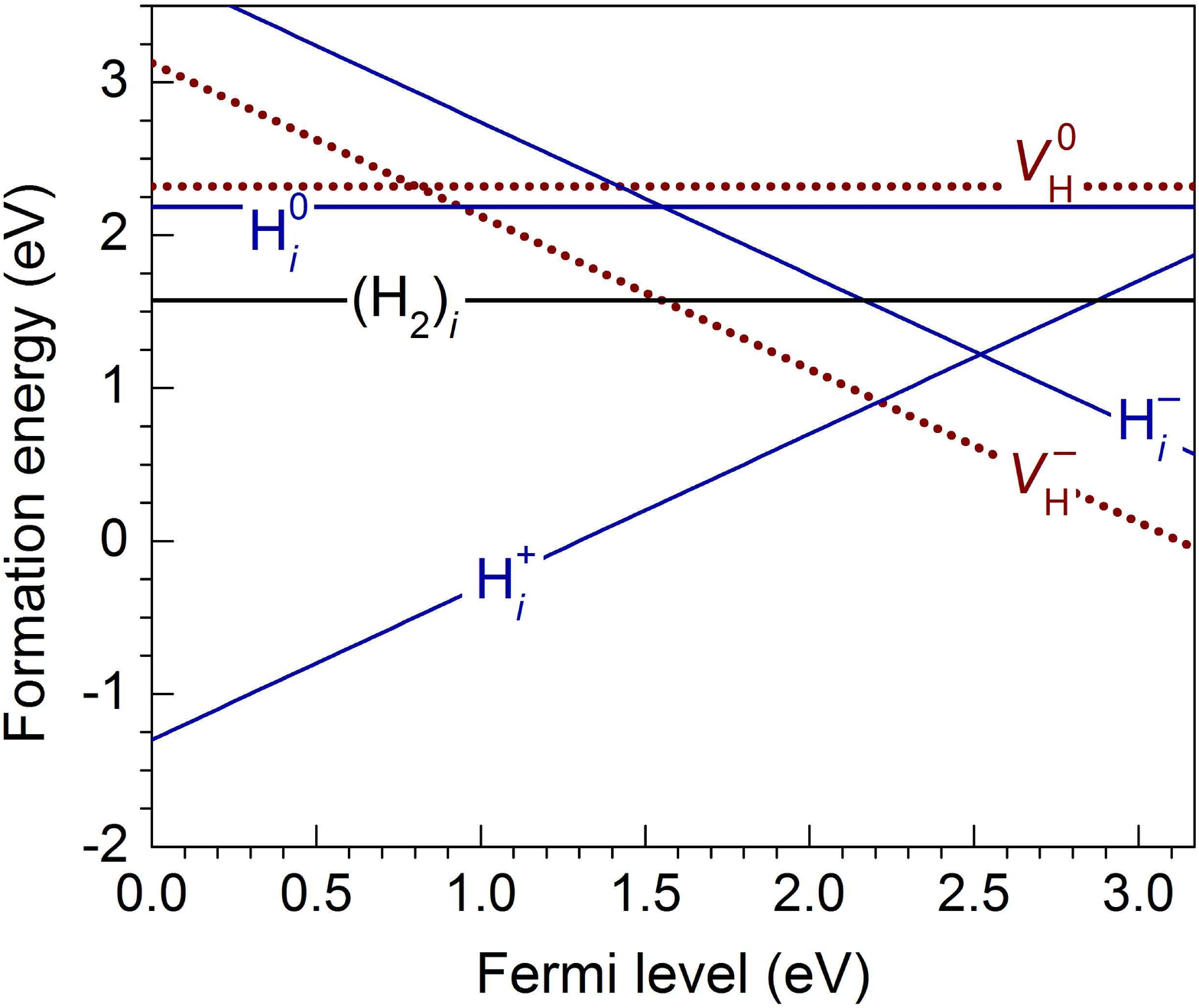}
\end{center}
\vspace{-0.15in}
\caption{(Color online) Calculated formation energies of hydrogen-related defects in LiNH$_{2}$, plotted as a function of Fermi energy with respect to the VBM.}\label{LiNH2;FE;H}
\end{figure}

\begin{figure*}
\begin{center}
\includegraphics[width=6.4in]{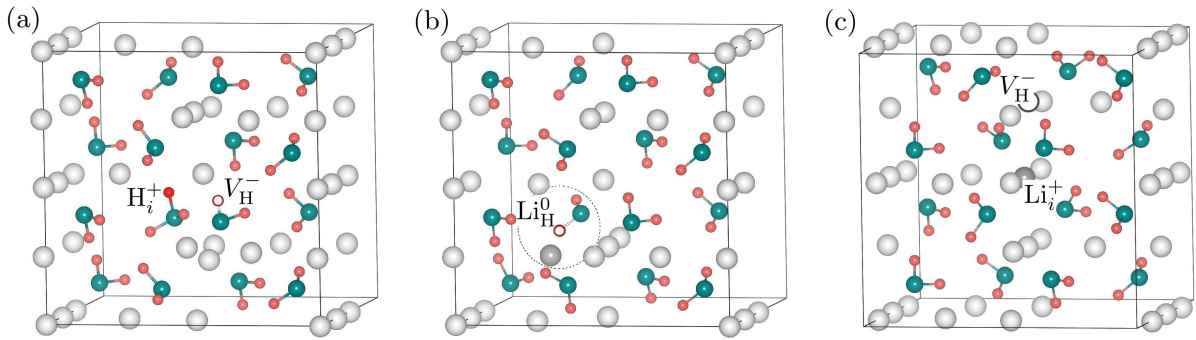}
\end{center}
\vspace{-0.15in}
\caption{(Color online) Structure of (a) (H$_{i}^{+}$,$V_{\mathrm{H}}^{-}$), (b) Li$_{\mathrm{H}}^{0}$, and (c) (Li$_{i}^{+}$,$V_{\mathrm{Li}}^{-}$) in LiNH$_{2}$. Large (gray) spheres are Li, medium (blue) spheres N, and small (red) spheres H. The vacancies are represented by an empty sphere.}\label{LiNH2;Struct}
\end{figure*}

Figure \ref{LiNH2;FE;H} shows the calculated formation energies for hydrogen vacancies ($V_{\mathrm{H}}$), interstitials (H$_{i}$), and interstitial molecules (H$_{2}$)$_{i}$ in LiNH$_{2}$. Among these native defects, the negatively charged hydrogen vacancy ($V_{\mathrm{H}}^{-}$) and positively charged hydrogen interstitial (H$_{i}^{+}$) have the lowest formation energies over the entire range of Fermi-level values. The neutral hydrogen vacancy ($V_{\mathrm{H}}^{0}$) and interstitial (H$_{i}^{0}$) are high in energy. The formation energy of (H$_{2}$)$_{i}$ is also higher than that of $V_{\mathrm{H}}^{-}$ and H$_{i}^{+}$. The positively charged hydrogen vacancy ($V_{\mathrm{H}}^{+}$, not included in Fig.~\ref{LiNH2;FE;H}) is unstable, i.e., a locally stable configuration of this defect cannot be stabilized. If we try to create $V_{\mathrm{H}}^{+}$, it decays to a situation where the positive charge is not associated with the point defect but corresponds to free carriers in the valence band.

In order to understand the energetics of different hydrogen-related defects in LiNH$_{2}$, it is useful to refer back to the electronic structure and bonding geometry of LiNH$_{2}$. For example, the creation of $V_{\mathrm{H}}$ involves breaking an N$-$H bond from the (NH$_2$)$^-$ unit, resulting in an NH unit. Since the NH unit is most favorable in the (NH)$^{2-}$ configuration due to the high electronegativity of the N atom, it is expected that $V_{\mathrm{H}}$ will be most stable in the $V_{\mathrm{H}}^-$ configuration. Formation of $V_{\mathrm{H}}^{0}$, on the other hand, would involve removing one electron from the resulting (NH)$^{2-}$ unit, which is energetically highly unfavorable. Figure \ref{LiNH2;FE;H} indeed shows $V_{\mathrm{H}}^-$ to be the most stable configuration.

The creation of H$_{i}^{0}$ or H$_{i}^{+}$ leads to the formation of an NH$_{3}$ unit, which is an (NH$_{2}$)$^-$ unit with an extra H atom. Since NH$_{3}$ forms a closed-shell unit, the interstitial is expected to be most stable in the H$_{i}^{+}$ configuration, in which the additional electron (which stabilized (NH$_{2}$)$^-$ but is now superfluous) is removed. H$_{i}^{-}$, on the other hand, prefers to stay in an interstitial void, with distances of 1.91 and 2.14 {\AA} to the two nearest Li atoms. Finally, the creation of (H$_{2}$)$_{i}$ involves adding an H$_{2}$ molecule to the system. This interstitial molecule prefers to stay near the center of the octahedron formed by six NH$_{2}$ units, with the calculated H$-$H bond length being 0.75 {\AA}, very close to that calculated for an isolated H$_{2}$ molecule.

For the migration of H$_{i}^{+}$, H$_{i}^{-}$, $V_{\rm{H}}^{-}$, and (H$_{2}$)$_{i}$, we find energy barriers of 0.61, 0.34, 0.71, and 0.19 eV, respectively. The energy barriers for H$_{i}^{+}$ and $V_{\rm{H}}^{-}$ are relatively high because the migration of these two defects involves breaking N$-$H bonds. For H$_{i}^{+}$, an H atom in the NH$_{3}$ unit moves to the nearest NH$_{2}$. The saddle-point configuration consists of an H atom located midway between two NH$_{2}$ units (i.e., NH$_{2}-$H$-$NH$_{2}$). Similarly, the migration of $V_{\rm{H}}^{-}$ involves moving an H atom from a nearby NH$_{2}$ unit to the vacancy. The saddle-point configuration in this case consists of a hydrogen atom located midway between two NH units (i.e., NH$-$H$-$NH). H$_{i}^{-}$ and (H$_{2}$)$_{i}$, on the other hand, can migrate without breaking and forming bonds, explaining their relatively low migration barriers. We note that the bond length of the H$_{2}$ dimer is preserved along the migration path of (H$_{2}$)$_{i}$.

We also investigated the formation of Frenkel pairs composed of H$_{i}$ and $V_{\mathrm{H}}$. The possible hydrogen-related Frenkel pairs are (H$_{i}^{+}$,$V_{\mathrm{H}}^{-}$) and (H$_{i}^{-}$,$V_{\mathrm{H}}^{+}$); the latter is not considered, since $V_{\mathrm{H}}^{+}$ is unstable. Figure~\ref{LiNH2;Struct}(a) shows the structure of (H$_{i}^{+}$,$V_{\mathrm{H}}^{-}$) in LiNH$_{2}$. The configurations of the individual defects are preserved in this complex; i.e., H$_{i}^{+}$ forms an NH$_{3}$ unit and the creation of $V_{\mathrm{H}}^{-}$ leaves an (NH)$^{2-}$ unit. The distance between the two N ions in the pair is 3.37 {\AA}, very close to the N$-$N distance in the bulk (3.38 {\AA}). This Frenkel pair has a formation energy of 1.54 eV, and a binding energy of 0.38 eV (with respect to the isolated constituents). We note that these quantities are independent of the choice of chemical potentials.

\begin{figure}
\begin{center}
\includegraphics[width=3.0in]{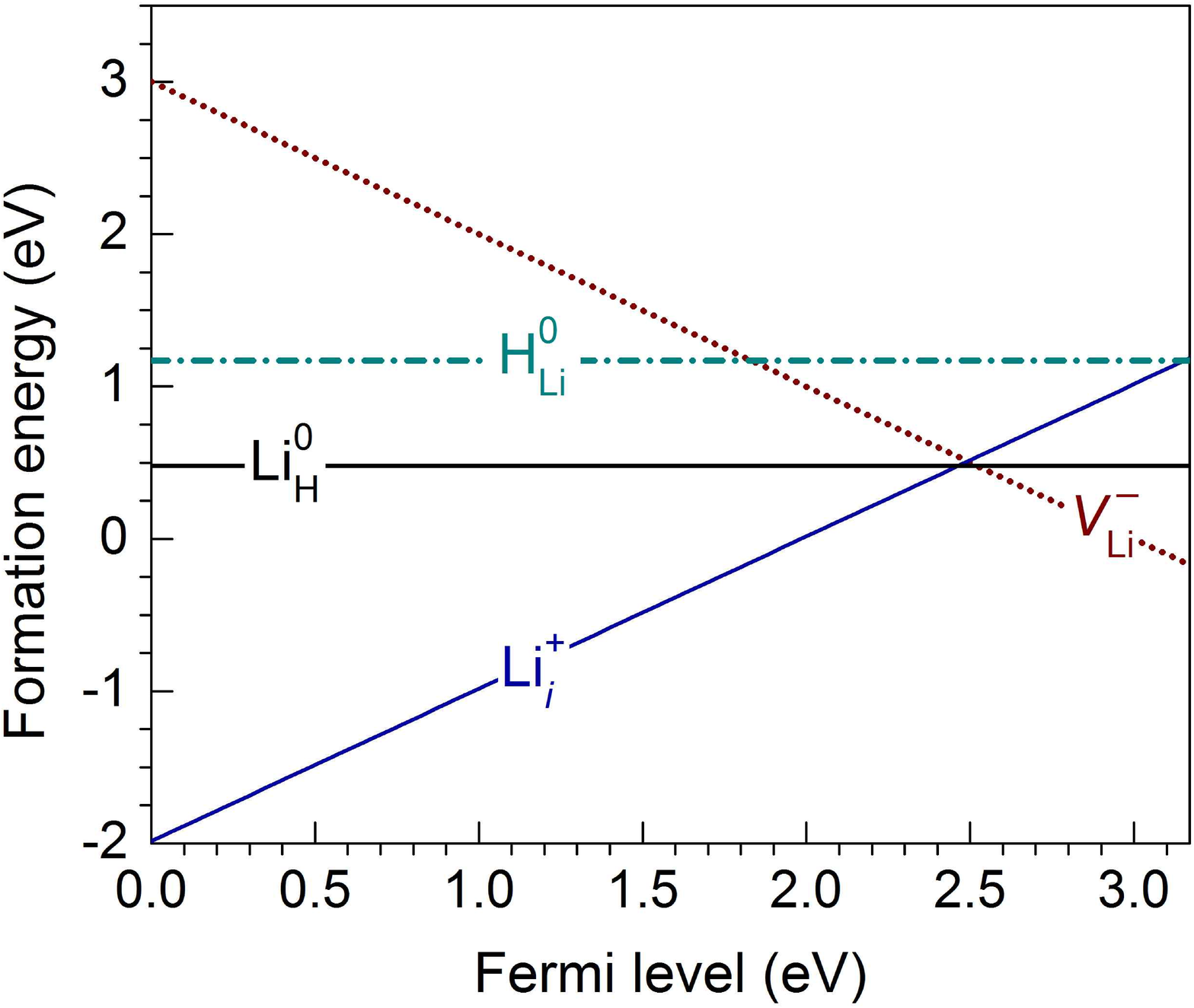}
\end{center}
\vspace{-0.15in}
\caption{(Color online) Calculated formation energies of lithium-related defects in LiNH$_{2}$, plotted as a function of Fermi energy with respect to the VBM.}\label{LiNH2;FE;Li}
\end{figure}

\subsubsection{Lithium-related defects}

Figure \ref{LiNH2;FE;Li} shows the calculated formation energies for lithium vacancies ($V_{\mathrm{Li}}$), interstitials (Li$_{i}$), Li$_{\mathrm{H}}^{0}$ (Li replacing an H atom), and H$_{\mathrm{Li}}^{0}$ (H replacing an Li atom) in LiNH$_{2}$. Among the lithium-related defects, Li$_{i}^{+}$ and $V_{\mathrm{Li}}^{-}$ have the lowest formation energies for all the Fermi-level values, except for a very small range near $\mu_{e}$=2.49 eV, where Li$_{\mathrm{H}}^{0}$ has a slightly lower formation energy. $V_{\mathrm{Li}}^{+}$ and Li$_{i}^{-}$ are unstable, $V_{\mathrm{Li}}^{0}$ and Li$_{i}^{0}$  and not shown in Fig.~\ref{LiNH2;FE;Li}.

In the case of $V_{\mathrm{Li}}^{-}$, a Li$^{+}$ ion was removed from the Li3 site ({\em cf.}~Fig.~\ref{structures}), whereas for Li$_{i}^{+}$, a Li$^{+}$ ion was placed in the void formed by two NH$_{2}$ units where one of the two N$-$H bonds in each NH$_{2}$ unit points toward the interstitial Li atom. We find that these defects lead to structural relaxations such that the neighboring Li atoms and NH$_{2}$ units are slightly displaced and rotated.

The formation of Li$_{\mathrm{H}}^{0}$, on the other hand, results in an NH unit and a Li atom in the nearby region; see Fig.~\ref{LiNH2;Struct}(b). Li$_{\mathrm{H}}^{0}$ can indeed be regarded as a complex of $V_{\mathrm{H}}^{-}$ and Li$_{i}^{+}$. The formation energy of Li$_{\mathrm{H}}^{0}$ is lower than the sum of the formation energies of Li$_{i}^{+}$ and $V_{\mathrm{H}}^{-}$ by 0.66 eV. In addition, considering the presence of the (NH)$^{2-}$ unit and the additional Li$^{+}$ ion, the region that includes Li$_{\mathrm{H}}^{0}$ can be locally considered as Li$_{2}$NH inside  bulk LiNH$_{2}$.

Finally, H$_{\mathrm{Li}}^{0}$ was created by replacing a Li atom with an H atom. This leaves the system with an NH$_{3}$ unit and a Li vacancy. H$_{\mathrm{Li}}^{0}$ can be regarded as a complex of H$_{i}^{+}$ and $V_{\mathrm{Li}}^{-}$ with a binding energy of 0.62 eV. Note that, if equilibrium between LiNH$_2$ and Li$_2$NH is assumed, the formation energies of Li$_{\mathrm{H}}^{0}$ and H$_{\mathrm{Li}}^{0}$ are independent of the chemical potentials because the chemical potential terms in their formation energies occur as ($-$$\mu_{\rm Li}$+$\mu_{\rm H}$), which is a constant, as seen from Eqs.~(\ref{eq:LiNH2}) and (\ref{eq:Li2NH}).

The migration of Li$_{i}^{+}$ involves moving the Li$^{+}$ ion between two ground-state configurations, giving an energy barrier as low as 0.30 eV. For $V_{\mathrm{Li}}^{-}$, the migration involves moving Li$^{+}$ from a nearby lattice site to the vacancy and this gives a barrier of 0.20 eV. These values are relatively small, suggesting that Li$_{i}^{+}$ and $V_{\mathrm{Li}}^{-}$ are highly mobile. For Li$_{\mathrm{H}}^{0}$, which is a complex of Li$_{i}^{+}$ and $V_{\mathrm{H}}^{-}$, a lower bound on the migration barrier is given by the migration barrier of the least mobile constituent,~\cite{wilsonshort09} i.e., 0.71 eV, the value for $V_{\mathrm{H}}^{-}$. Similarly, the migration barrier of H$_{\mathrm{Li}}^{0}$ is estimated to be 0.61 eV, the value for H$_{i}^{+}$.

We also investigated possible formation of lithium Frenkel pairs. Since Li$_{i}^{-}$ and $V_{\mathrm{Li}}^{+}$ are unstable, the only possibility is (Li$_{i}^{+}$,$V_{\mathrm{Li}}^{-}$), whose structure is shown in Fig.~\ref{LiNH2;Struct}(c). The distance between Li$_{i}^{+}$ and $V_{\mathrm{Li}}^{-}$ is 0.85 {\AA}. This pair has a formation energy of 0.65 eV and a binding energy of 0.36 eV. The formation energy is, therefore, much lower than that of the hydrogen Frenkel pair, i.e., (H$_{i}^{+}$,$V_{\mathrm{H}}^{-}$). This result indicates that LiNH$_{2}$ is likely to exhibit Frenkel disorder on the Li sublattice.

\begin{figure}
\begin{center}
\includegraphics[width=3.0in]{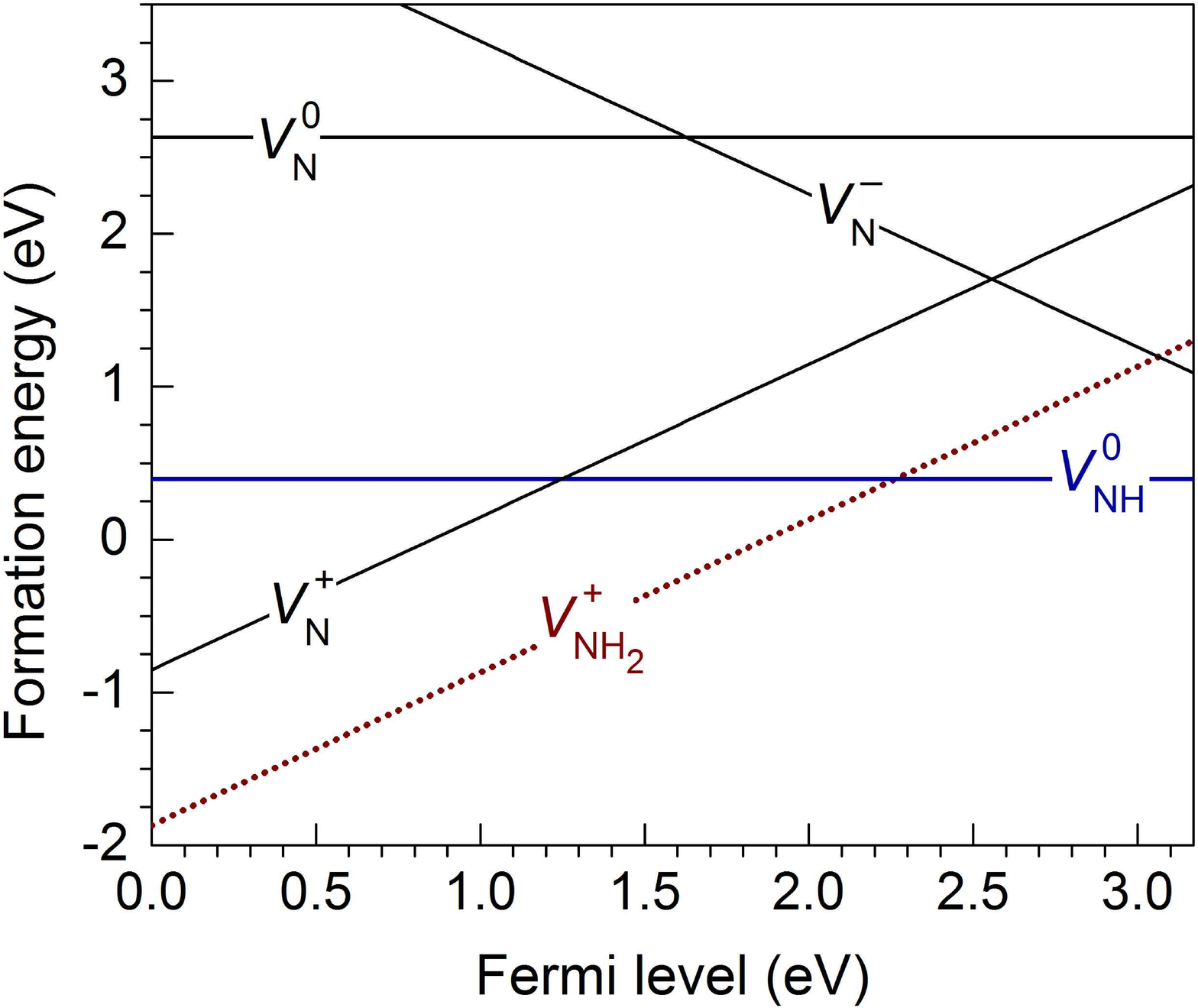}
\end{center}
\vspace{-0.15in}
\caption{(Color online) Calculated formation energies of nitrogen-related defects in LiNH$_{2}$, plotted as a function of Fermi energy with respect to the VBM.}\label{LiNH2;FE;N}
\end{figure}

\subsubsection{Nitrogen-related defects}
\label{ssec:N}

Figure \ref{LiNH2;FE;N} shows the calculated formation energies of nitrogen vacancies ($V_{\mathrm{N}}$), NH vacancies ($V_{\mathrm{NH}}$), and NH$_{2}$ vacancies ($V_{\mathrm{NH_{2}}}$) in LiNH$_{2}$. We find that $V_{\mathrm{NH_{2}}}$ is stable as $V_{\mathrm{NH_{2}}}^{+}$, and $V_{\mathrm{NH}}$ is stable in the neutral charge state ($V_{\mathrm{NH}}^0$). $V_{\mathrm{N}}$ is stable as $V_{\mathrm{N}}^+$ and $V_{\mathrm{N}}^-$. We also investigated interstitial NH$_{3}$ molecules but found them to have a very high formation energy (not included in Fig.~\ref{LiNH2;FE;N}), $E^{f}$=2.54 eV for the chosen set of chemical potentials. This suggests that ammonia is unlikely to form and diffuse through bulk LiNH$_{2}$ in the form of interstitial molecules.

$V_{\mathrm{NH_{2}}}^{+}$ corresponds to the removal of an entire (NH$_{2}$)$^{-}$ unit from bulk LiNH$_{2}$. We find that there is very little change in the local lattice structure surrounding this defect. The formation of $V_{\mathrm{NH}}^{0}$, on the other hand, leaves one H atom in the resulting void. This isolated H atom is surrounded by four Li atom with the Li$-$H distances in the range 1.95$-$2.15 {\AA}. $V_{\mathrm{NH}}^{0}$ can then be regarded as a complex of $V_{\mathrm{NH_{2}}}^{+}$ and H$_{i}^{-}$ with a binding energy of 1.56 eV. Similarly, $V_{\mathrm{N}}^{+}$ can be regarded as a complex composed of $V_{\mathrm{NH_{2}}}^{+}$ and (H$_{2}$)$_{i}$ with a binding energy of 0.74 eV; and $V_{\mathrm{N}}^{-}$ as a complex of $V_{\mathrm{NH_{2}}}^{+}$ and two H$_{i}^{-}$ defects with a binding energy of 1.53 eV.

The migration of $V_{\mathrm{NH_{2}}}^{+}$ involves moving a nearby (NH$_{2}$)$^{-}$ unit to the vacancy, with an energy barrier of 0.87 eV. For $V_{\mathrm{NH}}^{0}$, which can be considered as a complex of $V_{\mathrm{NH_{2}}}^{+}$ and H$_{i}^{-}$, a lower bound on the barrier is 0.87 eV, determined by the least mobile species, i.e., $V_{\mathrm{NH_{2}}}^{+}$.

Other groups have recently reported first-principles calculations for native defects in LiNH$_{2}$, using methodologies similar to ours.\cite{miceli,hazrati,wang} The calculated formation energies and migration barriers of individual hydrogen-, lithium-, and nitrogen-related defects reported by Wang {\it et al.}\cite{wang} are in close agreement with our results (to within 0.1 eV for most defects, with a maximum deviation of 0.2 eV in the case of $V_{\rm H}^-$, our value being lower). Comparing to the results of Hazrati {\it et al.},\cite{hazrati} the deviations are somewhat larger (up to 0.4 eV), for which we cannot offer an explanation. Hazrati {\it et al.}~did include vibrational zero-point energy corrections for those defects that involve hydrogen. However, while zero-point energies can be significant, a large degree of cancellation always occur between the terms in the solid and in the reservoirs and the effect on formation energies is typically small. Miceli {\it et al.}~did not report calculated formation energies of individual point defects. For the (H$_{i}^{+}$,$V_{\mathrm{H}}^{-}$) Frenkel pair, Hazrati {\it et al.}~and Wang {\it et al.}~reported formation energies of 1.66 eV and 1.93 eV, respectively, compared to 1.54 eV in our calculations. For the (Li$_{i}^{+}$,$V_{\mathrm{Li}}^{-}$) Frenkel pair, their reported values are 0.72 eV and 0.79 eV, whereas our calculated value is 0.65 eV. Miceli {\it et al.}, on the other hand, reported a formation energy of 0.97 eV for the lithium Frenkel pair. We attribute the differences in the results for the Frenkel pairs to differences in the atomic configuration of the pairs. Our lower energies indicate that the configurations we identified are more stable.

\subsection{Li$_{2}$NH}

\subsubsection{Hydrogen-related defects}

\begin{figure}
\begin{center}
\includegraphics[width=3.0in]{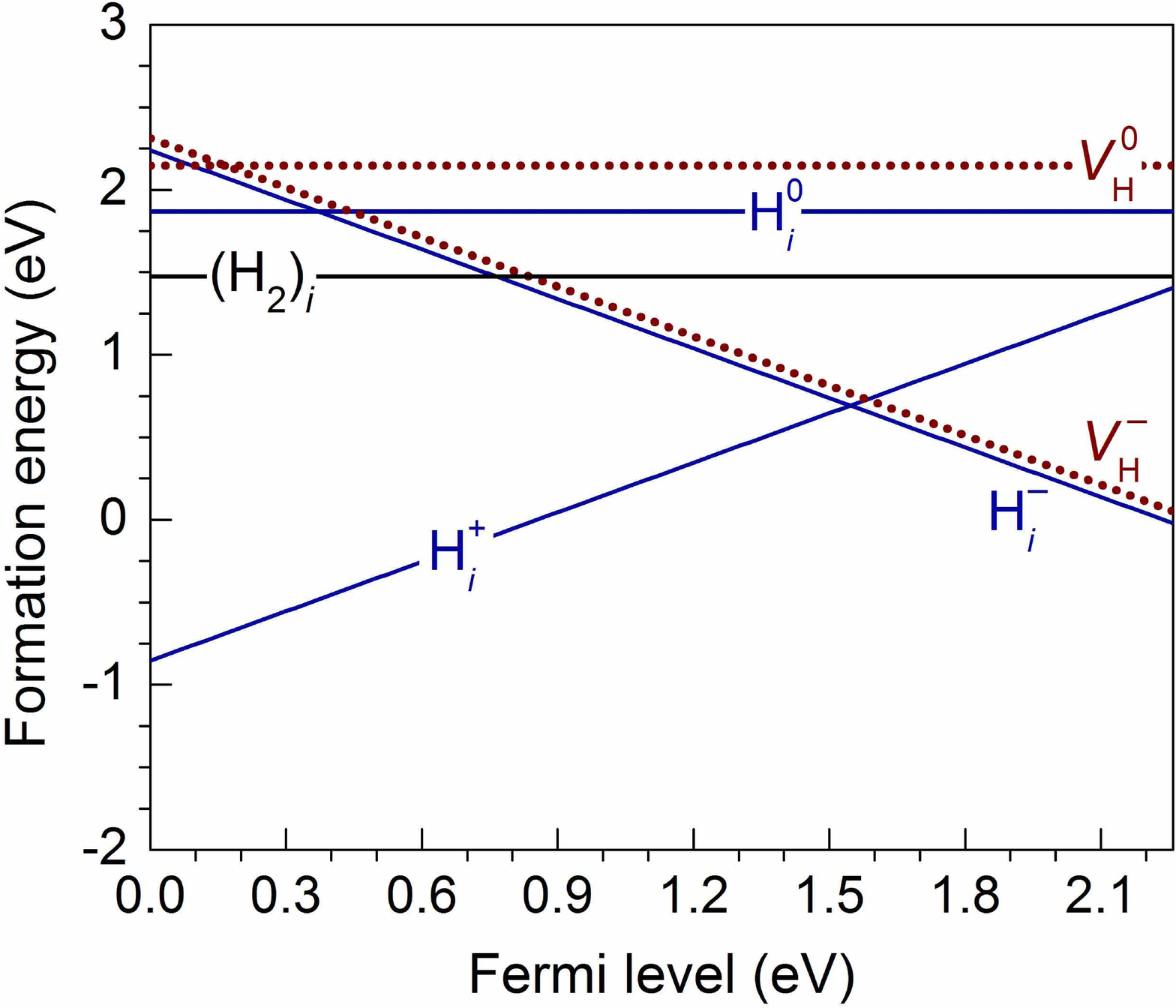}
\end{center}
\vspace{-0.15in}
\caption{(Color online) Calculated formation energies of hydrogen-related defects in Li$_{2}$NH, plotted as a function of Fermi energy with respect to the VBM.}\label{Li2NH;FE;H}
\end{figure}

\begin{figure*}
\begin{center}
\includegraphics[width=6.4in]{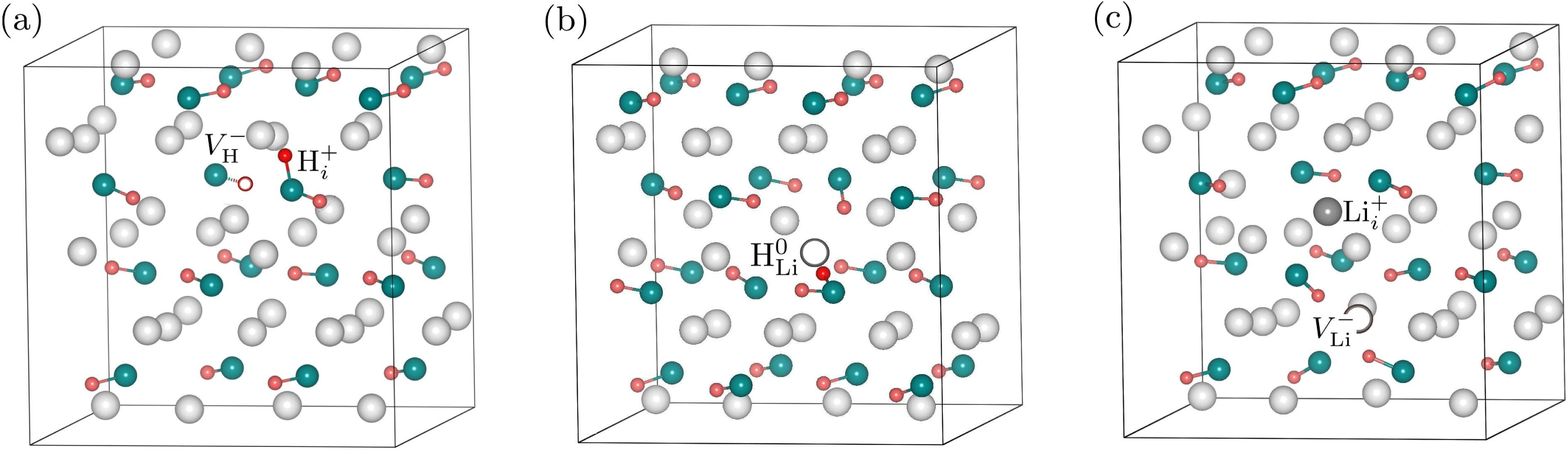}
\end{center}
\vspace{-0.15in}
\caption{(Color online) Structure of (a) (H$_{i}^{+}$,$V_{\mathrm{H}}^{-}$), (b) H$_{\mathrm{Li}}^{0}$, and (c) (Li$_{i}^{+}$,$V_{\mathrm{Li}}^{-}$) in Li$_{2}$NH. Large (gray) spheres are Li, medium (blue) spheres N, and small (red) spheres H. The vacancies are represented by an empty sphere.}\label{Li2NH;Struct}
\end{figure*}

Figure \ref{Li2NH;FE;H} shows the calculated formation energies for H$_{i}$, $V_{\mathrm{H}}$, and (H$_{2}$)$_{i}$ in Li$_{2}$NH. Among the hydrogen-related defects, H$_{i}^{+}$ and H$_{i}^{-}$ have the lowest formation energies for the chosen set of chemical potentials. Neutral defects such as $V_{\mathrm{H}}^{0}$ and H$_{i}^{0}$ are high in energy, and the formation energy of (H$_{2}$)$_{i}$ is also significantly higher than that of H$_{i}^{+}$ and H$_{i}^{-}$. The positively charged $V_{\mathrm{H}}^{+}$ is unstable.

In Li$_{2}$NH, the removal of one H atom from an (NH)$^{2-}$ unit to form $V_{\mathrm{H}}$ results in an isolated N atom. Since N has high electronegativity, it is expected that $V_{\mathrm{H}}$ would be most stable in the $V_{\mathrm{H}}^{-}$ configuration, consistent with our results shown in Fig.~\ref{Li2NH;FE;H}. The formation of H$_{i}^{+}$ results in an (NH$_{2}$)$^-$ unit. H$_{i}^{-}$, on the other hand, prefers to stay in an interstitial site near three Li atoms with the Li$-$H distances in the range 1.78$-$1.87 {\AA}. Finally, (H$_{2}$)$_{i}$ stays in an interstitial void, with a calculated H$-$H bond length of 0.77 {\AA}, comparable to but slightly larger than that calculated for an isolated H$_{2}$ molecule (0.75 {\AA}).

Regarding the migration of the hydrogen-related defects, we find energy barriers of 0.95, 0.65, and 1.66 eV for H$_{i}^{+}$, H$_{i}^{-}$, and $V_{\rm{H}}^{-}$, respectively. The migration barriers for H$_{i}^{+}$ and $V_{\rm{H}}^{-}$ are again high, even higher than in LiNH$_{2}$, because the migration of these two defects involves breaking of N$-$H bonds. For H$_{i}^{+}$, the H attached to an NH unit moves to the nearest NH unit. The saddle-point configuration consists of an H atom located midway between two NH units, i.e., NH$-$H$-$NH. Likewise, the migration of $V_{\rm{H}}^{-}$ involves moving an H$_{i}^{+}$ from an NH unit to the vacancy. The saddle-point configuration in this case consists of an H atom located midway between two N atoms, i.e., N$-$H$-$N.

Figure~\ref{Li2NH;Struct}(a) shows the structure of the (H$_{i}^{+}$,$V_{\mathrm{H}}^{-}$) Frenkel pair in Li$_{2}$NH. Similar to the (H$_{i}^{+}$,$V_{\mathrm{H}}^{-}$) pair in LiNH$_{2}$, the configurations of individual defects are also preserved in this complex; i.e., H$_{i}^{+}$ forms an NH$_{2}$ unit and $V_{\mathrm{H}}^{-}$ leaves the system with an isolated N atom. The distance between the two N atoms in the pair is 3.39 {\AA}, comparable to the N$-$N distance in the bulk (3.31 {\AA}). (H$_{i}^{+}$,$V_{\mathrm{H}}^{-}$) has a formation energy of 1.32 eV and a binding energy of 0.14 eV. This low value of the binding energy suggests that, once created, the pair will easily dissociate.

\begin{figure}
\begin{center}
\includegraphics[width=3.0in]{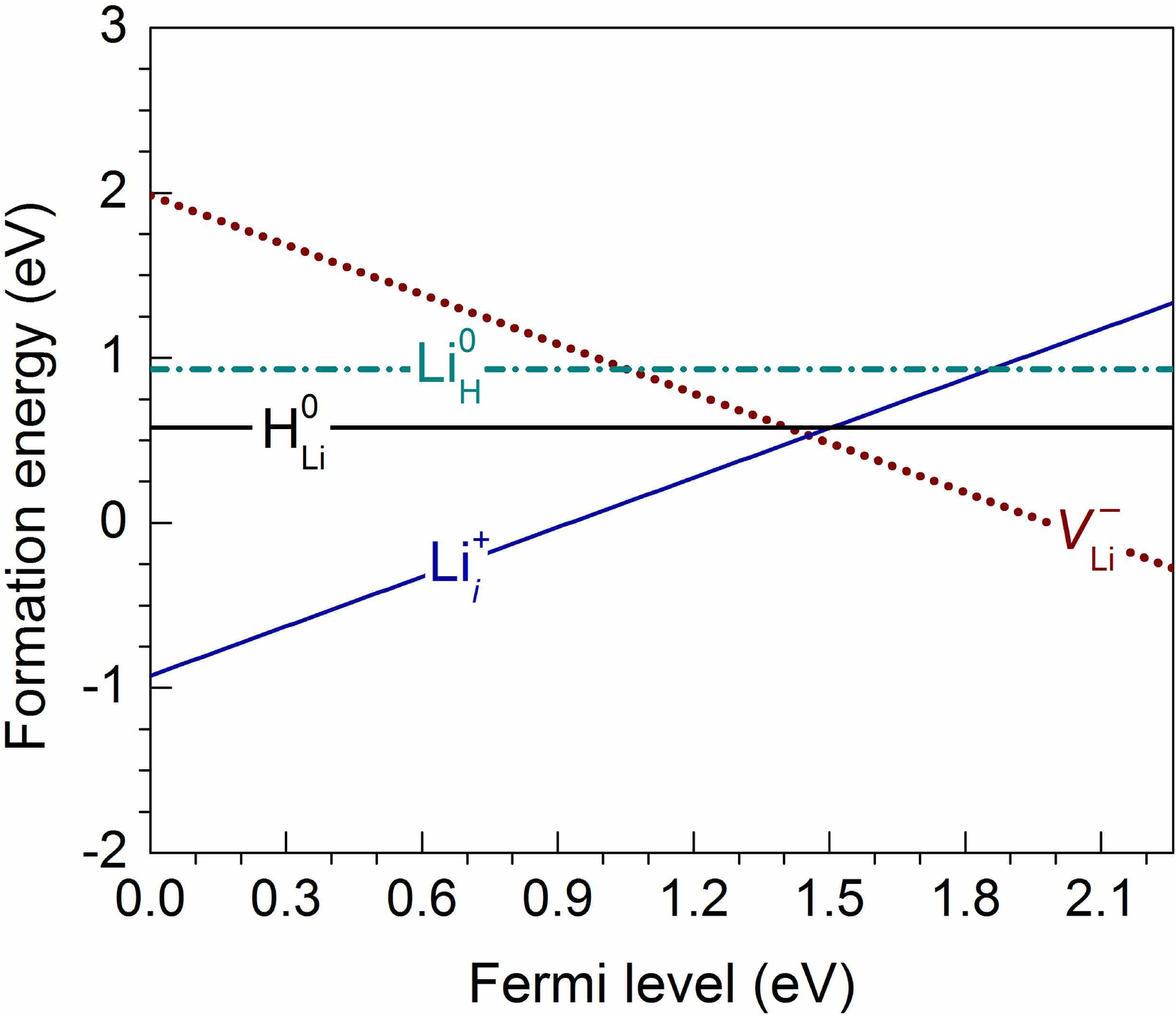}
\end{center}
\vspace{-0.15in}
\caption{(Color online) Calculated formation energies of lithium-related defects in Li$_{2}$NH, plotted as a function of Fermi energy with respect to the VBM.}\label{Li2NH;FE;Li}
\end{figure}

\subsubsection{Lithium-related defects}

Figure \ref{Li2NH;FE;Li} shows the calculated formation energies for $V_{\mathrm{Li}}$, Li$_{i}$, H$_{\mathrm{Li}}^{0}$ (H replacing a Li atom), and Li$_{\mathrm{H}}^{0}$ (Li replacing a H atom) in Li$_{2}$NH. Among these defects, Li$_{i}^{+}$ and $V_{\mathrm{Li}}^{-}$ have the lowest formation energies. H$_{\mathrm{Li}}^{0}$ also has a relatively low formation energy. $V_{\mathrm{Li}}^{+}$, $V_{\mathrm{Li}}^{0}$, Li$_{i}^{-}$, and Li$_{i}^{0}$ are unstable. Note that, if equilibrium between LiNH$_2$ and Li$_2$NH is assumed, the formation energies of H$_{\mathrm{Li}}^{0}$ and Li$_{\mathrm{H}}^{0}$ are independent of the chemical potentials, similar to the equivalent defects in LiNH$_{2}$.

$V_{\mathrm{Li}}^{-}$ in Li$_2$NH corresponds to the removal of a (Li)$^{+}$ unit from the system, whereas Li$_{i}^{+}$ can be thought of as the addition of a Li$^{+}$ ion to the system. These two defects result in relatively small local perturbations in the Li$_2$NH lattice. The creation of H$_{\mathrm{Li}}^{0}$, on the other hand, leaves the system with an NH$_{2}$ unit and a Li vacancy, as seen in Fig.~\ref{Li2NH;Struct}(b). Thus, H$_{\mathrm{Li}}^{0}$ can be regarded as a complex of H$_{i}^{+}$ and $V_{\mathrm{Li}}^{-}$. The formation energy of H$_{\mathrm{Li}}^{0}$ is lower than the sum of the formation energies of H$_{i}^{+}$ and $V_{\mathrm{Li}}^{-}$ by 0.55 eV. Since the resulting defects are an NH$_{2}$ unit and a Li vacancy, the region that includes H$_{\mathrm{Li}}^{0}$ can be considered as locally LiNH$_{2}$ inside bulk Li$_{2}$NH.

Finally, Li$_{\mathrm{H}}^{0}$ was created by replacing an H atom with a Li atom. This results in an N atom standing near seven Li atoms with Li$-$N distances of less than 2.2 {\AA}. Li$_{\mathrm{H}}^{0}$ can actually be considered as a complex of Li$_{i}^{+}$ and $V_{\mathrm{H}}^{-}$ with a binding energy of 0.45 eV. This defect can act as a nucleation site for Li$_{3}$N formation in the dehydrogenation reaction of Li$_{2}$NH. For comparison, the Li$-$N bonds are 1.94 and 2.11 {\AA} in bulk Li$_{3}$N.

The migration barriers of Li$_{i}^{+}$ and $V_{\mathrm{Li}}^{-}$ are 0.29 and 0.14 eV, respectively. For H$_{\mathrm{Li}}^{0}$, which is a complex of H$_{i}^{+}$ and $V_{\mathrm{Li}}^{-}$, we estimate a migration barrier of 0.95 eV, the value for H$_{i}^{+}$. Similarly, the migration barrier of Li$_{\mathrm{H}}^{0}$ is estimated to be 1.66 eV, the value for $V_{\mathrm{H}}^{-}$.

Figure~\ref{Li2NH;Struct}(c) shows the structure of the (Li$_{i}^{+}$,$V_{\mathrm{Li}}^{-}$) Frenkel pair in Li$_{2}$NH. The distance between Li$_{i}^{+}$ and $V_{\mathrm{Li}}^{-}$ is 3.13 {\AA}. The (Li$_{i}^{+}$,$V_{\mathrm{Li}}^{-}$) pair has a formation energy of 0.68 eV and a binding energy of 0.38 eV. The formation energy is much lower than that of the (H$_{i}^{+}$,$V_{\mathrm{H}}^{-}$) pair. This suggests that Li$_{2}$NH, like LiNH$_{2}$, is also prone to Frenkel disorder on the Li sublattice.

\begin{figure}
\begin{center}
\includegraphics[width=3.4in]{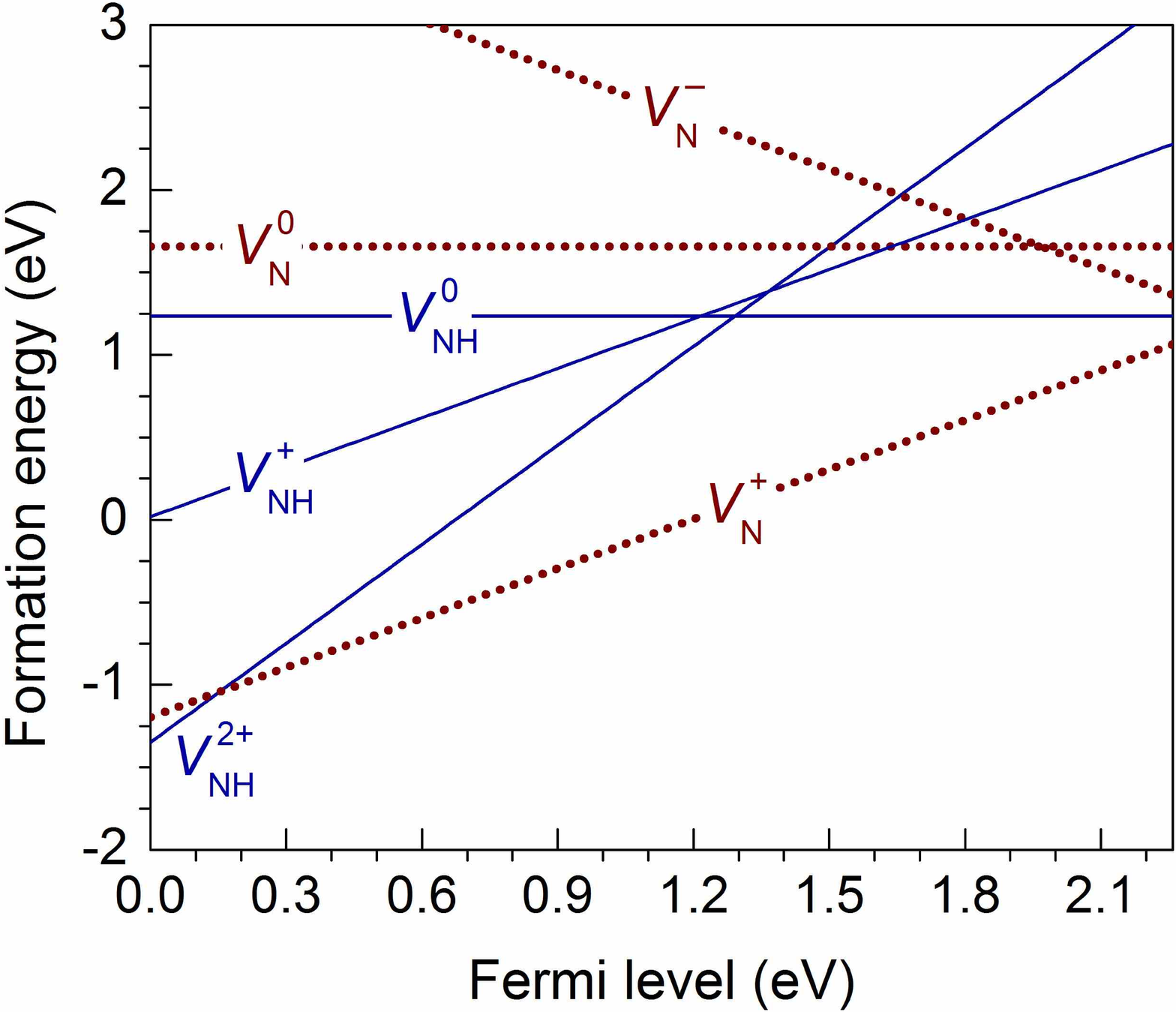}
\end{center}
\vspace{-0.15in}
\caption{(Color online) Calculated formation energies of nitrogen-related defects in Li$_{2}$NH, plotted as a function of Fermi energy with respect to the VBM.}\label{Li2NH;FE;N}
\end{figure}

\subsubsection{Nitrogen-related defects}

Figure \ref{Li2NH;FE;N} shows the calculated formation energies for $V_{\mathrm{N}}$ and $V_{\mathrm{NH}}$ in Li$_{2}$NH. Of all the possible nitrogen-related defects, $V_{\mathrm{N}}^{+}$ has the lowest formation energy for almost all Fermi-level values. $V_{\mathrm{N}}^{+}$ can be regarded as a complex of $V_{\mathrm{NH}}^{2+}$ and H$_{i}^{-}$, with a binding energy of 2.09 eV. The isolated H atom (i.e., H$_{i}^{-}$) is surrounded by six Li atom with the Li$-$H distances in the range 2.00$-$2.36 {\AA}. $V_{\mathrm{N}}^{0}$ and $V_{\mathrm{N}}^{-}$ have high formation energies and are thus unlikely to form.

$V_{\mathrm{NH}}^{2+}$ in Li$_{2}$NH is similar to $V_{\mathrm{NH_{2}}}^{+}$ in LiNH$_{2}$, meaning they are both created by removing an entire anionic unit, i.e., (NH$_{2}$)$^{-}$ or (NH)$^{2-}$, from the bulk compounds. But, unlike $V_{\mathrm{NH_{2}}}^{+}$ in LiNH$_{2}$, which was stable over a wide range of Fermi levels (see Fig.~\ref{LiNH2;FE;N}), $V_{\mathrm{NH}}^{2+}$ in Li$_{2}$NH is stable only over a very narrow range of Fermi levels
near the VBM (Fig.~\ref{Li2NH;FE;N}). Likewise, $V_{\mathrm{N}}^{+}$ in Li$_{2}$NH is similar to $V_{\mathrm{NH}}^{0}$ in LiNH$_{2}$ because they both have a H$_{i}^{-}$ in the interstitial void formed by removing an anionic unit.

For the migration of $V_{\mathrm{NH}}^{2+}$ in Li$_{2}$NH, we find an energy barrier of 0.91 eV. For $V_{\mathrm{N}}^{+}$, the estimated energy barrier is also 0.91 eV, the energy barrier for $V_{\mathrm{NH}}^{2+}$.

We have also investigated interstitial NH$_{3}$ molecules in Li$_{2}$NH and find that they have relatively high formation energies if the NH$_{3}$ unit is preserved. Instead, we find that the NH$_{3}$ molecule prefers to combine with a host (NH)$^{2-}$ unit to form two (NH$_{2}$)$^-$ units, lowering the energy by 0.54 eV. Even with this lower-energy configuration, the formation energy of 2.60 eV is still too high for it to be a relevant defect. Our results clearly indicate that NH$_{3}$ is unlikely to form and diffuse as interstitial molecules in bulk Li$_{2}$NH (as we already found in the case of LiNH$_{2}$).

\section{Discussion}

\begin{table}
\caption{Calculated formation energies ($E^{f}$) and migration barriers ($E_{m}$) for native defects in LiNH$_2$ and Li$_{2}$NH. Atomic chemical potentials were chosen to reflect equilibrium with LiNH$_{2}$ and Li$_2$NH, and the experimental conditions at which the (de)hydrogenation processes occur (see text). Migration energies denoted by an asterisk ($^{\ast}$) are estimated by considering the defect as a complex (last column in the Table) and taking the higher of the migration energies of the constituents.}\label{tab}
\begin{center}
\begin{ruledtabular}
\begin{tabular}{cllll}
&Defect&$E^f$ (eV)& $E_m$ (eV)&Complex \\
%&&&&			\\
\colrule
LiNH$_{2}$
&H$_{i}^+$ &1.28&0.61&	\\
&H$_{i}^-$ &1.34&0.34&	\\
&$V_{\rm{H}}^-$		&0.63&0.71&	\\
&(H$_{2}$)$_{i}$	&1.75&0.19&	\\
&Li$_{i}^+$		 &0.51&0.30&	\\
&$V_{\rm{Li}}^-$	&0.51&0.20&	\\
&Li$_{\rm{H}}^0$	&0.48&0.71$^{\ast}$&Li$_{i}^{+}$+$V_{\rm{H}}^{-}$	\\
&H$_{\rm{Li}}^0$	&1.17&0.61$^{\ast}$&H$_{i}^{+}$+$V_{\rm{Li}}^{-}$	\\
&$V_{\rm{NH_2}}^+$	&0.62&0.87	\\
&$V_{\rm{NH}}^0$	&0.40&0.87$^{\ast}$&$V_{\rm{NH}_{2}}^{+}$+H$_{i}^{-}$	\\
&$V_{\rm{N}}^+$	 &1.64&0.87$^{\ast}$&$V_{\rm{NH}_{2}}^{+}$+(H$_{2}$)$_{i}$	\\
&$V_{\rm{N}}^-$	 &1.77&0.87$^{\ast}$&$V_{\rm{NH}_{2}}^{+}$+2H$_{i}^{-}$	\\
\colrule
Li$_{2}$NH
&H$_{i}^+$ &0.74&0.95&	\\
&H$_{i}^-$ &0.65&0.65&	\\
&$V_{\rm{H}}^-$		&0.72&1.66&	\\
&(H$_{2}$)$_{i}$	&1.47&-&	\\
&Li$_{i}^+$		 &0.66&0.30&	\\
&$V_{\rm{Li}}^-$	&0.39&0.14&	\\
&H$_{\rm{Li}}^0$	&0.58&0.95$^{\ast}$&H$_{i}^{+}$+$V_{\rm{Li}}^{-}$\\
&Li$_{\rm{H}}^0$	&0.93&1.66$^{\ast}$&Li$_{i}^{+}$+$V_{\rm{H}}^{-}$\\
&$V_{\rm{NH}}^{2+}$	&1.83&0.91&	\\
&$V_{\rm{N}}^+$	 &0.39&0.91$^{\ast}$&$V_{\rm{NH}}^{2+}$+H$_{i}^{-}$	\\
\end{tabular}
\end{ruledtabular}
\end{center}
\end{table}

Table~\ref{tab} lists formation energies and migration barriers for all relevant native defects in LiNH$_{2}$ and Li$_2$NH. For charged defects in LiNH$_{2}$, we set $\mu_{e}$=2.49 eV, where the formation energies of Li$_{i}^{+}$ and $V_{\rm{Li}}^{-}$ are equal. This choice of Fermi level is based on the assumption that electrically active impurities are either absent or present in lower concentrations than the native point defects. In this case, the Fermi level is determined by oppositely charged defects with lowest formation energies, i.e., Li$_{i}^{+}$ and $V_{\rm{Li}}^{-}$ for the chosen set of chemical potentials in LiNH$_{2}$ that represents the dehydrogenation conditions ($\mu_{\rm H}$=$-$0.49 eV). The charge neutrality condition then requires these defects to be present in equal concentrations.\cite{peles07,wilsonshort09,hoang2009} Similarly, in the case of Li$_2$NH the defect formation energies are taken at $\mu_{e}$=1.59 eV, i.e., the Fermi level value at which the formation energies of $V_{\mathrm{N}}^{+}$ and $V_{\mathrm{Li}}^{-}$ are equal, where the chemical potentials are chosen to represent the hydrogenation conditions ($\mu_{\rm H}$=$-$0.31 eV).

It emerges from our analysis in the previous sections that the structure and energetics of all relevant native defects in LiNH$_{2}$ and Li$_{2}$NH can be interpreted in terms of basic building blocks, which include H$_{i}^+$, H$_{i}^-$, $V_{\rm{H}}^-$, (H$_{2})_{i}$, Li$_{i}^+$, $V_{\rm{Li}}^-$, and $V_{\rm{NH_2}}^+$ (or $V_{\rm{NH}}^{2+}$). Understanding the electronic and structural properties of these elementary defects is, therefore, crucial for understanding the defect complexes and the role these defects play in mass transport and ionic conduction. Based on the results presented in Sec.~IV, in the following we discuss Li-ion conduction in LiNH$_{2}$ and Li$_{2}$NH, and propose mechanisms for the decomposition of LiNH$_2$ and hydrogenation of Li$_2$NH. We also discuss the dehydrogenation of LiNH$_{2}$+LiH mixtures and the effects of ball milling.

\subsection{Li-ion conduction}

Let us first discuss ionic mobility on the Li sublattice and its consequences for ionic conduction. It is evident from Table~\ref{tab} that, in both LiNH$_{2}$ and Li$_2$NH, Li$_{i}^{+}$ and $V_{\rm{Li}}^{-}$ have low formation energies and are highly mobile. The (Li$_{i}^{+}$,$V_{\rm{Li}}^{-}$) pair that is composed of these two defects also has a low formation energy, 0.65 eV in LiNH$_{2}$ and 0.68 eV in Li$_2$NH, suggesting that Li$_{i}^{+}$ and $V_{\rm{Li}}^{-}$ can be created in the interior of the materials via a Frenkel pair mechanism. Our results are therefore in agreement with recent studies by Ludue\~{n}a {\it et al.}~using first-principles path integral molecular dynamics simulations and solid-state $^1$H NMR experiments where they observed significant disorder on the Li sublattice.~\cite{luduena}

Experimentally, Li$_{2}$NH was found to be a good ionic conductor, with an activation energy of 0.58 eV.~\cite{boukamp} This conductivity has been ascribed to the high mobility of Li ions. Our calculations show that both Li$_{i}^{+}$ and $V_{\rm{Li}}^{-}$ can contribute to the ionic conductivity. However, since the calculated migration barrier of $V_{\rm{Li}}^{-}$ is lower than that of Li$_{i}^{+}$, we expect that in Li$_{2}$NH (and LiNH$_{2}$) lithium diffusion by the vacancy mechanism is dominant. The calculated activation energy for self-diffusion of $V_{\rm{Li}}^{-}$ in Li$_{2}$NH is estimated to be 0.53 eV (the formation energy plus the migration barrier, {\it cf.}~Table~\ref{tab}), which is very close to the experimental activation energy.~\cite{boukamp} Similarly, we estimate the activation energy for self-diffusion of $V_{\rm{Li}}^{-}$ in LiNH$_{2}$ to be 0.71 eV, somewhat lower than the reported experimental value (0.94 eV).~\cite{matsuo,matsuo2011} As discussed in the next sections, the highly mobile Li$_{i}^{+}$ and $V_{\rm{Li}}^{-}$ also play an important role in the decomposition of LiNH$_{2}$ and hydrogenation of Li$_{2}$NH.

\subsection{Decomposition of LiNH$_{2}$}
\label{ssec:decomp}

Here we address the decomposition of LiNH$_{2}$ into Li$_{2}$NH and NH$_{3}$ according to reaction (\ref{eq:reaction3}). The transformation from LiNH$_{2}$ to Li$_{2}$NH involves breaking N$-$H bonds. This can be accomplished through the formation of $V_{\mathrm{H}}^{-}$, which in turn can occur in the interior of the material or at the surface. The required energies are not necessarily the same. The creation of $V_{\mathrm{H}}^{-}$ in the interior of LiNH$_2$, for instance, is necessarily accompanied by the creation of H$_{i}^{+}$ so that mass and charge are conserved. At the surface, one can create $V_{\mathrm{H}}^{-}$ by removing a proton (H$^{+}$) from LiNH$_{2}$ and this H$^{+}$ could be accommodated as an adsorbed atom or react with nearby species. These two possibilities, namely forming $V_{\mathrm{H}}^{-}$ in the interior of LiNH$_2$ or at the surface, can be interpreted as two different possible mechanisms for the reaction. As discussed below, the difference in the activation energies of these two mechanisms will lead to an effective dependence on the surface-to-volume ratio or the specific surface area (SSA) which can be measured experimentally. First we describe the mechanisms in more detail:

{\it Mechanism 1}: $V_{\mathrm{H}}^{-}$ and H$_{i}^{+}$ are created simultaneously in the interior of LiNH$_{2}$ through forming a (H$_{i}^{+}$,$V_{\mathrm{H}}^{-}$) Frenkel pair, i.e., moving H$^{+}$ from a lattice site to an interstitial site. This results in an (NH)$^{2-}$ next to an NH$_3$ unit representing $V_{\mathrm{H}}^{-}$ and H$_{i}^{+}$, respectively, as shown in Fig.~\ref{LiNH2;Struct}(a). Next, $V_{\mathrm{H}}^{-}$ and H$_{i}^{+}$ become separated as H$_{i}^{+}$ jumps from one (NH$_2$)$^-$ unit to another. This is equivalent to displacing the NH$_3$ unit away from the (NH)$^{2-}$ unit, leaving two Li$^+$ next to (NH)$^{2-}$; i.e., a formula unit of Li$_2$NH is locally formed inside LiNH$_2$. H$_{i}^{+}$ then migrates to the surface and is released as NH$_{3}$. Note that here we assume that as H$_{i}^{+}$ migrates from one (NH$_{2}$)$^{-}$ unit to the next, a corresponding Li$^+$ moves in the opposite direction in the form of Li$_{i}^+$ (see more below). The overall activation energy ($E_{a}$) for this mechanism then consists of the formation energy of the (H$_{i}^{+}$,$V_{\mathrm{H}}^{-}$) Frenkel pair (1.54 eV), the cost for separating the species in this Frenkel pair (0.38 eV), plus the migration barrier of H$_{i}^{+}$ (0.61 eV), i.e., $E_{a}$=1.54+0.38+0.61=2.53 eV. This activation energy is in very good agreement with the experimental value of 2.53 eV for the activation energy related to the decomposition of LiNH$_{2}$ before ball milling.~\cite{markmaitree}

{\it Mechanism 2}: $V_{\rm H}^-$ is created at the surface by removing an H$^{+}$ from LiNH$_{2}$. This H$^{+}$ ion can combine with a surface (NH$_2$)$^-$ unit to form NH$_3$ that is subsequently released. Given the ionic nature of the bonding between Li$^+$ and (NH$_2$)$^-$, we believe that such a process will be possible, irrespective of the details of the surface structure.  Note that the rate-limiting step in this mechanism is not the formation of $V_{\rm H}^-$ at the surface, but the hydrogen mass transport to the surface; i.e., in order to maintain this reaction, hydrogen atoms have to be transported to the surface. Here our only assumption is that the formation energy of $V_{\rm H}^-$ on the surface is lower than (or equal to) the formation energy in the bulk, which is a safe assumption given that the bonding environment at the surface is less constrained than in the bulk. In this mechanism, the activation energy is given by hydrogen self-diffusion mediated by $V_{\rm H}^-$, i.e., the sum of its formation energy and migration barrier: $E_{a}$=0.63+0.71=1.34 eV. The Li$^+$ unit that was left with after the surface (NH$_2$)$^-$ unit was released with the H$^{+}$ (in form of NH$_3$) assists the hydrogen self-diffusion in the form of Li$_i^+$, as required by the charge neutrality condition. Note also that the complex formed by $V_{\rm H}^-$ and Li$_{i}^+$ corresponds to a formula unit of Li$_2$NH inside LiNH$_2$. The calculated activation energy of 1.34 eV is also in good agreement with experimentally determined activation energies for the decomposition of ball-milled LiNH$_{2}$, ranging from 1.33 to 1.43 eV.~\cite{pinkerton05,markmaitree}

Since Mechanism 1 starts with the formation of defects in the bulk and Mechanism 2 with the formation of defects at the surface, we expect the prevalent mechanism and hence the effective activation energy for decomposition to be dependent on the surface-to-volume ratio. In samples composed of sufficiently large particles of LiNH$_2$, the surface-to-volume ratio is small and Mechanism 1 prevails. On the other hand, in samples composed of relatively small particles, i.e., with large surface-to-volume ratio, Mechanism 2 prevails. Indeed, it has been observed that in LiNH$_2$ samples subjected to ball milling, the activation energy for decomposition decreases with milling time, from 2.53 eV (before ball milling, SSA: 3.72 m$^{2}$/g) to 2.30 eV (after 45min of milling, SSA: 40.71 m$^{2}$/g) to 1.43 eV (after 3h, SSA: 46.65 m$^{2}$/g);\cite{markmaitree} i.e., as the milling time increases the particle size is decreased and the SSA increased, and we expect the prevalent mechanism to change from 1 to 2. These experimental activation energy values are within the range (1.34$-$2.52 eV) established by the calculated activation energies for Mechanisms 1 and 2. It should be noted that the increase in SSA upon ball milling not only increases the likelihood of point defect formation at the surface, it also increases the chance that the point defects can reach all parts of the ``bulk'' within a given amount of time. While surfaces are of course present even in Mechanism 1, they simply fail to make enough of a difference to modify the observed activation energy.

In both mechanisms the highly mobile and low-formation-energy Li$_{i}^{+}$ and $V_{\rm{Li}}^{-}$ provide local charge neutrality and additional mass transport. Without the accompanying Li$_{i}^{+}$ defect, for example, $V_{\mathrm{H}}^{-}$ would not be able to diffuse into the bulk because local charge neutrality has to be maintained. On the other hand, Li$_{\mathrm{H}}^{0}$ (a complex of Li$_{i}^+$ and $V_{\rm H}^-$) in LiNH$_{2}$ and H$_{\mathrm{Li}}^{0}$ (a complex of H$_{i}^+$ and $V_{\rm Li}^-$) in Li$_2$NH have very low formation energies, suggesting that Li amide (imide) can be locally formed within the bulk Li imide (amide). Our results therefore support David {\it et al.}'s observations that the Li amide/imide reaction is a bulk reaction, and that there is a continuous transformation between LiNH$_{2}$ and Li$_2$NH via non-stoichiometric intermediates.~\cite{davidJACS}

We acknowledge that Mechanisms 1 and 2, which are based on calculations of point defects in the dilute limit, do not present a complete picture of the decomposition process. However, the formation and migration of point defects is an initial, but essential and critical, step toward decomposition. In this initial step, the concentration of point defects will be low, thus justifying our focus on the dilute limit. Other processes certainly play a role as well in the ultimate decomposition, but the agreement with experiment indicates that these other processes have activation energies that are either lower than, or comparable to, the point-defect-related mechanisms we are describing. In addition, the fact that we predict different activation energies for different particle sizes, in agreement with experiment, provides support for the point-defect mechanisms indeed being the rate-limiting step.

As mentioned in Sec.~\ref{sec:intro}, other research groups have also tried to understand the decomposition of LiNH$_{2}$ into Li$_{2}$NH and NH$_{3}$ based on first-principles defect calculations. Although not clearly stated, Miceli {\it et al.}\cite{miceli} seemed to suggest that for small LiNH$_{2}$ particles the decomposition process occurs at the surface with the formation of (H$_{i}^{+}$,$V_{\mathrm{H}}^{-}$) Frenkel pairs; and for larger particles, the formation of (H$_{i}^{+}$,$V_{\mathrm{H}}^{-}$) would also occur in the bulk. This is somewhat similar to the two mechanisms we described above. However, Miceli {\it et al.}~suggested further that, in the former case, the rate-limiting step at the early stage of decomposition is the formation of (H$_{i}^{+}$,$V_{\mathrm{H}}^{-}$) at the surface in the presence of lithium Frenkel pairs. This is different from our Mechanism 2 where the rate-limiting step is self-diffusion of $V_{\mathrm{H}}^{-}$. Hazrati {\it et al.}\cite{hazrati} also proposed that the decomposition process occurs at the surface with the formation of (H$_{i}^{+}$,$V_{\mathrm{H}}^{-}$) Frenkel pairs. Wang {\it et al.},\cite{wang} on the other hand, did not provide any specific mechanism but suggested that the formation of H$_{i}^{+}$ is the rate-limiting step in hydrogen mass transport.

\subsection{Dehydrogenation of LiNH$_{2}$+LiH mixtures}
\label{ssec:dehydro}

The mechanisms we have proposed can also provide an understanding of the dehydrogenation of LiNH$_{2}$+LiH mixtures, i.e., reaction (\ref{eq:reaction2}). In these systems, one expects that LiNH$_{2}$ and LiH are in intimate contact if the reactants are carefully mixed. At the LiNH$_{2}$/LiH interface, LiH can provide H$^-$ ions.  Our calculated formation energy for $V_{\rm H}^+$ vacancies in LiH is 0.69 eV, and since indiffusion of $V_{\rm H}^+$ is equivalent to outdiffusion of H$_i^-$, this result confirms that LiH can indeed supply the H$^-$ ions that we invoke. These H$^-$ ions can combine with H$_{i}^{+}$ (that is created in the bulk of LiNH$_{2}$ and migrates to the LiNH$_{2}$/LiH interface via Mechanism 1) or H$^{+}$ (that is liberated from LiNH$_{2}$ when creating $V_{\mathrm{H}}^{-}$ via Mechanism 2) to form H$_{2}$ without releasing any NH$_{3}$.  This explains the formation of H$_{2}$ in reaction (\ref{eq:reaction2}).
If LiNH$_{2}$ and LiH are not in intimate contact, NH$_{3}$ can still be produced from LiNH$_{2}$ according to reaction (\ref{eq:reaction3}) because the H$^{-}$ (from LiH) is not immediately available to combine with H$_{i}^{+}$ or H$^{+}$ before the latter is released from LiNH$_{2}$ in the form of NH$_{3}$. In this case, the resulting NH$_{3}$ can be captured by LiH according to reaction (\ref{eq:reaction4}) and/or released as one of the products.

It has been demonstrated that the activation energy for the dehydrogenation of LiNH$_2$+LiH mixtures also decreases with increasing ball-milling time.~\cite{shaw2008,varin} Shaw {\it et al.}~reported  activation energies of 1.70 eV (SSA: 4.65 m$^{2}$/g), 1.36 eV (SSA: 47.36 m$^{2}$/g), 1.18 eV (SSA: 51.32 m$^{2}$/g), and 0.65 eV (SSA: 62.35 m$^{2}$/g) for the dehydrogenation of the LiNH$_2$+LiH mixture before ball milling and after the samples were ball-milled for 1.5h, 3h, and 24h, respectively.~\cite{shaw2008} Varin {\it et al.}, on the other hand, reported a different set of activation energies: 2.46 eV (before milling, SSA: 16.5 m$^{2}$/g), 0.98 eV (after 1h, SSA: 26.4 m$^{2}$/g), 0.88 eV (after 25h, SSA: 59.6 m$^{2}$/g), and 0.91 eV (after 100h, SSA: 45.6 m$^{2}$/g).~\cite{varin} Both sets of experimental values show the same trend: the  activation energy is reduced significantly with ball milling and there is a correlation with the measured SSA.

We suggest that the activation energy for the dehydrogenation of LiNH$_2$+LiH mixtures with relatively short milling times is predominantly determined by that for the decomposition of LiNH$_{2}$. The above mentioned experimental data can therefore be explained in terms of our discussion in Sec.~\ref{ssec:decomp} about LiNH$_{2}$ decomposition, meaning the dehydrogenation of the mixtures is expected to proceed via Mechanisms 1 and/or 2, and the extent to which one mechanism dominates over the other depends on the surface-to-volume ratio (or the SSA). This provides an explanation for the observed activation energies in the range from 1.34 to 2.52 eV. For those samples that exhibit activation energies lower than that of Mechanism 2 (1.34 eV), produced after long milling times, we suggest that the milling process may have created a high degree of damage in the LiNH$_2$+LiH mixtures, even to the point of local amorphization. Formation energies for defects in these damaged regions would be lower than in the pristine bulk, resulting in defect concentrations well above the equilibrium concentrations; this lowering of the cost of forming the rate-limiting defects results in a lowering of the activation energy for dehydrogenation.

Shaw {\it et al.}~suggested that NH$_{3}$ diffusion through a Li$_{2}$NH product layer outside a LiNH$_{2}$ shrinking core is the rate-limiting step in the kinetics of the dehydrogenation of LiNH$_2$+LiH mixtures.~\cite{markmaitree,shawJPS} We find that this is very unlikely if the Li$_{2}$NH layer is thick enough. As presented in Sec.~\ref{sec:defects}, our results clearly indicate that NH$_{3}$ is not likely to form (and diffuse) as interstitial molecules in either LiNH$_{2}$ or Li$_{2}$NH because the formation energy is too high. In Li$_{2}$NH, interstitial NH$_{3}$ molecules are even unstable toward forming (NH$_{2}$)$^-$ units, by combining with host (NH)$^{2-}$ units.

Note that the calculated activation energy of Mechanism 2 reported in Sec.~\ref{ssec:decomp} depends on the formation energy of $V_{\mathrm{H}}^{-}$ at the Fermi-level value $\mu_{e}$ determined by the charge neutrality condition, which in turn depends on the chemical potentials of Li, N, and H. However, we have checked several possible scenarios and found that the calculated activation energy is not sensitive to the choice of chemical potentials. In the case of LiNH$_2$+LiH mixtures, for example, if the two reactants are carefully mixed, one can assume equilibrium between LiNH$_{2}$, Li$_{2}$NH, and LiH, which gives rise to a different set of chemical potentials where $\mu_{\mathrm{H}}$=$-$0.40 eV. The Fermi level of LiNH$_2$ is then at $\mu_{e}$=2.58 eV where Li$_{i}^{+}$ and $V_{\mathrm{Li}}^{-}$ have equal formation energies. We find that in this case the activation energy of Mechanism 2 is still 1.34 eV.

\subsection{Hydrogenation of Li$_{2}$NH}

Before discussing the hydrogenation mechanism of Li$_{2}$NH, let us summarize what is known about the hydrogenation process in metals. The absorption of hydrogen to form a metal hydride includes several steps:~\cite{vincent} (i) the applied H$_{2}$ is physisorbed on the surface of the metal; (ii) the physisorbed H$_{2}$ is dissociated at the surface and becomes chemisorbed; (iii) H atoms move to subsurface sites and diffuse through the metal; (iv) as the hydrogen concentration increases, a metal hydride phase nucleates. In this process, the rate-limiting step changes from the dissociation and penetration of hydrogen at the metal/H$_{2}$ interface to the nucleation of the hydride phase, and possibly the diffusion of hydrogen through the metal hydride layer that forms around the metal particle.~\cite{vincent} We expect to see similar processes in Li$_{2}$NH.

For the hydrogenation reaction in Eq.~(\ref{eq:reaction2}), the highly mobile Li$_{i}^{+}$ and $V_{\mathrm{Li}}^{-}$ in Li$_{2}$NH are expected to play an important role. These two defects can be created at the surface or simultaneously in the interior of the material via a Frenkel pair mechanism. Li$_{i}^{+}$ is likely to interact with the applied H$_{2}$ gas at the surface, or with the chemisorbed H that diffuses into the material, and form LiH and H$_{i}^{+}$, i.e., Li$_{i}^{+}$ + H$_{2}$ $\rightarrow$ LiH + H$_{i}^{+}$. This H$_{i}^{+}$ will then be attracted toward the $V_{\mathrm{Li}}^{-}$ defect to form H$_{\mathrm{Li}}^{0}$ (a complex of H$_{i}^{+}$ and $V_{\mathrm{Li}}^{-}$), which provides (NH$_{2}$)$^{+}$ units for the formation of LiNH$_{2}$. This is similar to the mechanism proposed by David {\it et al.}\cite{davidJACS} for Li amide/imide hydrogenation. The rate-limiting step in this process could be the diffusion of H$_{i}^{+}$ in the bulk of Li$_{2}$NH. However this cannot be claimed with certainty without explicit investigations of all other possible steps involved in the hydrogenation process.

\section{Summary}

We have carried out comprehensive first-principles studies of native defects in LiNH$_{2}$ and Li$_{2}$NH. Both compounds are found to be prone to Frenkel disorder on the Li sublattice, which is consistent with experimental observations. Lithium interstitials and vacancies have low formation energies and are highly mobile; they can therefore participate in ionic conduction and mass transport, and act as accompanying defects for hydrogen-related defects in mass transport. Hydrogen interstitials and vacancies, on the other hand, are responsible for forming and breaking N$-$H bonds, which are essential in the Li amide/imide reaction. Based on the structure, energetics, and migration of hydrogen-, lithium-, and nitrogen-related point defects and defect complexes, we have proposed that LiNH$_{2}$ decomposes into Li$_{2}$NH and NH$_{3}$ according to two competing mechanisms, one involving the formation of native defects in the interior of the material, and the other at the surface. As a result, the prevalent mechanism and hence the effective activation energy for decomposition depend on the surface-to-volume ratio or the specific surface area, which changes with particle size during ball milling. These mechanisms also provide an explanation for the particle-size dependence of the activation energy of the decomposition of LiNH$_{2}$ and that of the dehydrogenation of LiNH$_{2}$+LiH mixtures.

\begin{acknowledgments}

K.H.~was supported by General Motors Corporation, and A.J.~by the U.S.~Department of Energy (Grant No.~DE-FG02-07ER46434). We acknowledge the use of the CNSI Computing Facility under NSF Grant No.~CHE-0321368, NERSC resources supported by the DOE Office of Science under Contract No.~DE-AC02-05CH11231, and the Ranger supercomputer from the TeraGrid computing resources supported by the NSF under Grant No.~DMR070072N.

\end{acknowledgments}

%\bibliography{references}% Produces the bibliography via BibTeX.

\end{document}